\documentclass[11pt]{article}
\pdfoutput=1

\usepackage{amsfonts}
\usepackage{amscd}
\usepackage{amssymb}
\usepackage{amsmath,bbm}
\usepackage{graphicx}
\usepackage{epsfig}
\usepackage{latexsym}
\usepackage{mathtools}
\usepackage[table]{xcolor}
\usepackage[vcentermath]{youngtab}
\usepackage{jheppub}
\usepackage{bbold}
\usepackage{subcaption}    
    
\def \be  {\begin{equation}}
\def \ee  {\end{equation}}
\def \bea {\begin{equation}\begin{aligned}}
\def \eea {\end{aligned}\end{equation}}
\def \ba  {\begin{eqnarray}}
\def \ea  {\end{eqnarray}}
\def \bb  {\begin{thebibliography}}
\def \eb  {\end{thebibliography}}
\def \lab #1 {\label{#1}}

\newcommand\cI{\mathcal{I}}
\newcommand\cJ{\mathcal{J}}

\newcommand\cL{\mathcal{L}}
\newcommand\cM{\mathcal{M}}
\newcommand\cN{\mathcal{N}}
\newcommand\cO{\mathcal{O}}

\newcommand\cV{\mathcal{V}}

\newcommand\cZ{\mathcal{Z}}
\newcommand\al{\alpha}

\newcommand\C{\mathbb{C}}
\newcommand\ep{\epsilon}

\newcommand\R{\mathbb{R}}

\newcommand\Q{{\bf Q}}

\newcommand\la{\langle}
\newcommand\ra{\rangle}

\newcommand\tr{\mathrm{Tr}}

\newcommand\vacman{\mathfrak{V}}

\allowdisplaybreaks
\graphicspath{{./Figures/}}

\title{Expanding the Bethe/Gauge Dictionary}

\author[1]{Mathew Bullimore,}\emailAdd{mathew.bullimore@maths.ox.ac.uk}
\author[2]{Hee-Cheol Kim,}\emailAdd{heecheol1@gmail.com}
\author[1]{Tomasz \L ukowski}\emailAdd{lukowski@maths.ox.ac.uk}

\affiliation[1]{Mathematical Institute, University of Oxford, Andrew Wiles Building, Radcliffe Observatory Quarter, Woodstock Road, Oxford, OX2 6GG, UK}
\affiliation[2]{Jefferson Physical Laboratory, Harvard University, Cambridge, MA 02138, USA}

\abstract{We expand the Bethe/Gauge dictionary between the XXX Heisenberg spin chain and 2d $\cN = (2,2)$ supersymmetric gauge theories to include aspects of the algebraic Bethe ansatz. We construct the wave functions of off-shell Bethe states as orbifold defects in the A-twisted supersymmetric gauge theory and  study their correlation functions. We also present an alternative description of off-shell Bethe states as boundary conditions in an effective $\cN = 4$ supersymmetric quantum mechanics. Finally, we interpret spin chain R-matrices as correlation functions of Janus interfaces for mass parameters in the supersymmetric quantum mechanics.}

\begin{document}
\setcounter{tocdepth}{2}
\maketitle

\section{Introduction}
\label{sec:intro}

\begin{figure}[htp]
\centering
\includegraphics[height=3.5cm]{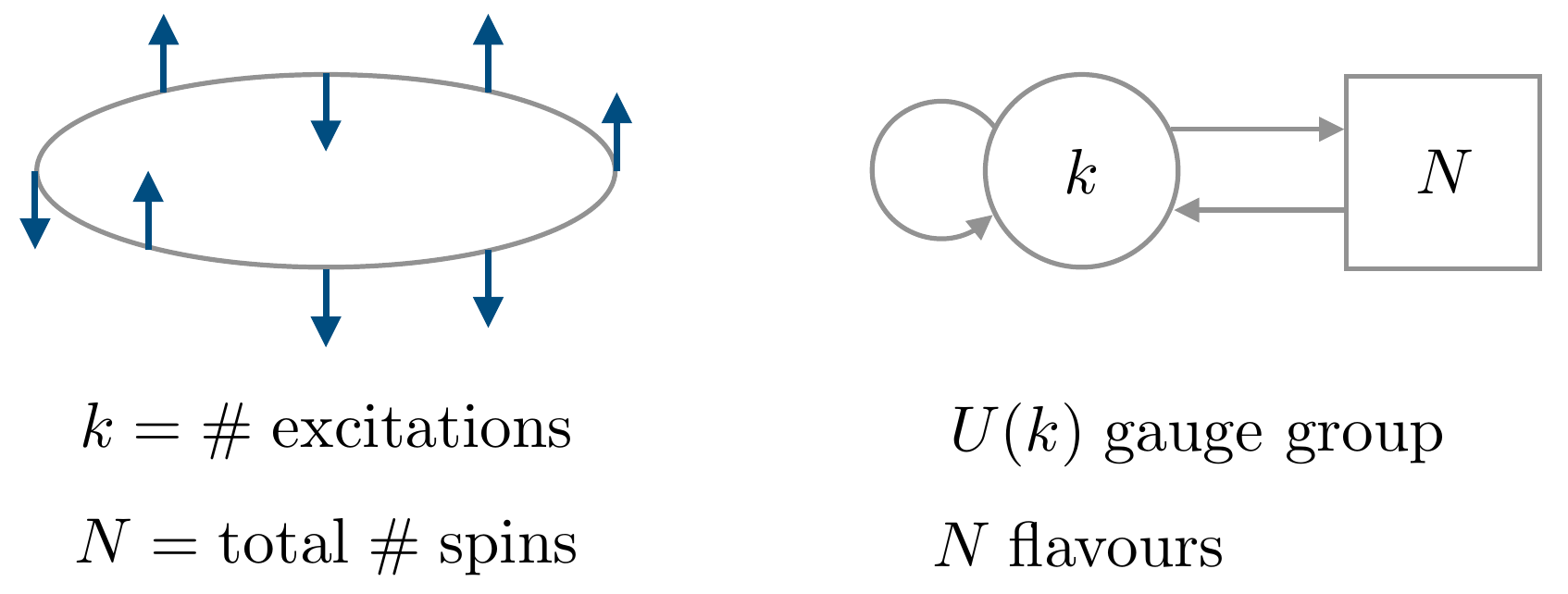}
\caption{Basic aspects of the correspondence between spin chains and supersymmetric gauge theory.}
\label{fig:intro}
\end{figure}

The aim of this paper is to extend the dictionary between quantum integrable systems and supersymmetric gauge theories introduced and studied in \cite{Nekrasov:2009uh,Nekrasov:2009ui,Nekrasov:2009rc}, the so-called Bethe/Gauge correspondence. We focus on an elementary example of this phenomenon: the correspondence between the XXX$_{\frac{1}{2}}$ Heisenberg spin chain and a family of 2d $\cN=(2,2)$ supersymmetric gauge theories. Some basic aspects of this correspondence are summarized in figure~\ref{fig:intro}.

A fundamental entry in the dictionary is the identification of the eigenstates of the spin chain Hamiltonian with massive vacua of the supersymmetric gauge theory. The spin chain is a quantum integrable system, whose spectrum is encoded in the Bethe equations~\cite{Bethe1931},
\be
\label{eq:Bethe.eqn}
\prod_{i=1}^N\frac{\sigma_a-m_i+\frac{\hbar}{2}}{\sigma_a-m_i-\frac{\hbar}{2}} =q\,\prod_{b\neq a}\frac{\sigma_a-\sigma_b+\hbar}{\sigma_a-\sigma_b-\hbar}\,, \qquad a=1,\ldots ,k.
\ee
Here, $\sigma_a$ denote the rapidities of excitations, $m_i$ are inhomogeneities, $q$ determines a quasi-periodic boundary condition and $\hbar$ is Planck's constant.

On the supersymmetric side of the correspondence, $m_j$ and $\hbar$ are complex mass parameters associated to flavour symmetries of the gauge theory in figure~\ref{fig:intro}, while $q$ is the exponential of a complexified FI parameter. For generic values of the mass parameters, the theory has a low energy description as a $U(1)^k$ gauge theory with an effective twisted superpotential $\widetilde W(\sigma)$ depending on complex vectormultiplet scalars $\sigma_a$. This function is determined exactly by a one-loop calculation. The equations for supersymmetric vacua,
\be
\label{eq:twisted-chiral-ring}
\exp{\frac{\partial \widetilde W}{\partial \sigma_a} } = 1\,, \qquad a = 1,\ldots,k\, ,
\ee
coincide with the Bethe equations for the spin chain~\eqref{eq:Bethe.eqn}. The ring generated by gauge-invariant functions of the vectormultiplet scalars $\sigma_1,\ldots,\sigma_k$ modulo the relations~\eqref{eq:twisted-chiral-ring} is the twisted chiral ring of the supersymmetric gauge theory. The twisted superpotential itself can be identified with Yang-Yang function of the spin chain.

A powerful approach to computing a wide range of observables in quantum integrable systems is the algebraic Bethe ansatz, as explained in~\cite{Faddeev:1996iy}. In this paper, we will understand how elements of this approach arise in the Bethe/Gauge correspondence. For this purpose, we will perform exact computations in the original supersymmetric gauge theory shown in figure~\ref{fig:intro}, rather than the effective abelian description. In particular, we will interpret aspects of the algebraic Bethe ansatz in terms of correlation functions in the $A$-type topological twist of the supersymmetric gauge theory, using techniques from supersymmetric localization~\cite{Closset:2015rna,Benini:2015noa}. Investigations of the Bethe/Gauge correspondence in this context have appeared in~\cite{Nekrasov:2014xaa,Chung:2016lrm}. The remainder of the introduction is dedicated to summarizing our results.

\begin{figure}[htp]
\centering
\includegraphics[height=3.5cm]{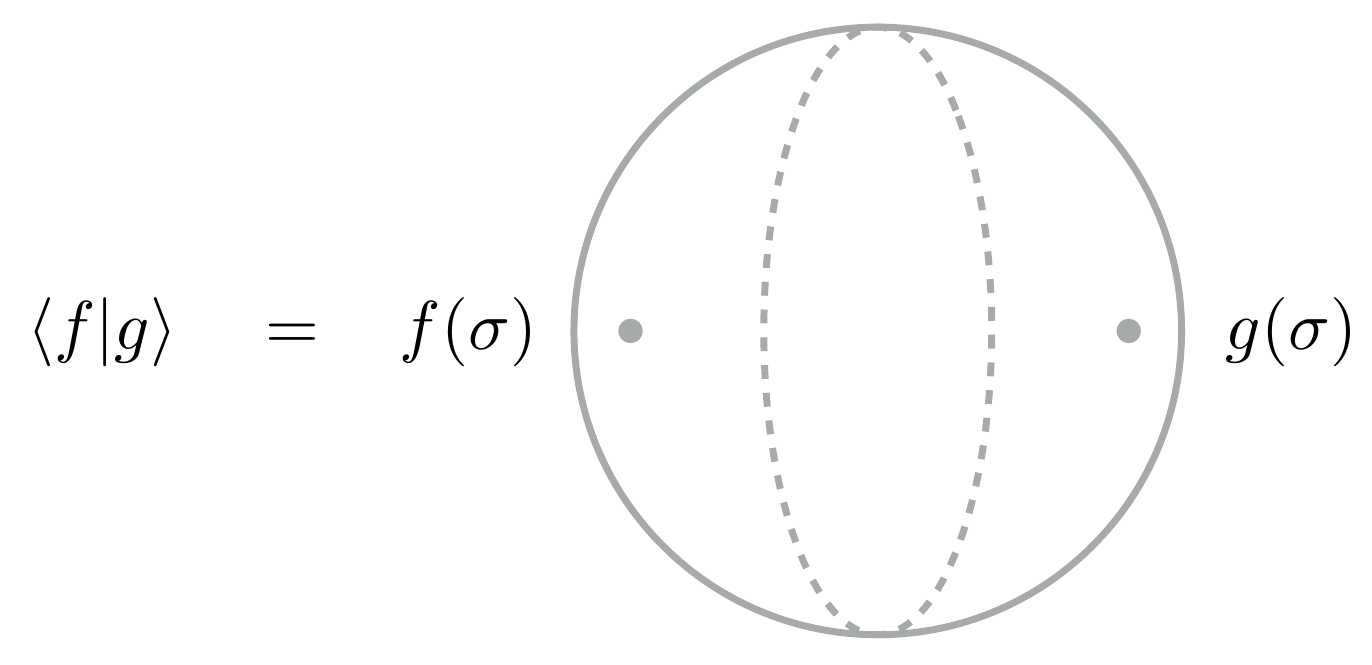}
\caption{The inner product on the spin chain Hilbert space corresponds to two-point correlation functions in the A-twisted supersymmetric gauge theory on $\mathbb{CP}^1$.}
\label{fig:intro-2pt}
\end{figure}

An important part of the algebraic Bethe ansatz is the construction of off-shell Bethe states $|\sigma_1,\ldots,\sigma_k\ra$, which are elements of the spin chain Hilbert space depending on auxiliary parameters $\sigma_1\ldots,\sigma_k$. The inner product $\la f | \sigma_1,\ldots,\sigma_k\ra$ with another state $|f\ra$ is a symmetric function $f(\sigma_1,\ldots,\sigma_k)$ of the auxiliary parameters, which can be identified with a gauge-invariant function of the vectormultiplet scalar in the supersymmetric gauge theory in figure~\ref{fig:intro}. The correlation functions of such operators in the A-type topological twist depend only on the class $[f(\sigma_1,\ldots,\sigma_k)]$ of the function modulo the twisted chiral ring relations~\eqref{eq:twisted-chiral-ring}. The map
\be
|f\ra \to [ f(\sigma_1,\ldots,\sigma_k) ]\,,
\ee
then sets up a correspondence between states in the spin chain Hilbert space and invariant functions of $\sigma_1,\ldots,\sigma_a$ modulo relations, such that the inner product $\la f | g\ra$ on the spin chain Hilbert space coincides with the two-point correlation function of $f(\sigma_1,\ldots,\sigma_k)$ and $g(\sigma_1,\ldots,\sigma_k)$ in the A-twisted theory on $\mathbb{CP}^1$. This is illustrated in figure~\ref{fig:intro-2pt}.

In order to investigate this relation, it is convenient to introduce an orthonormal `up-down' basis for the spin chain Hilbert space. The basis elements are labelled by subsets $I = \{i_1,\ldots,i_k\}  \subset \{1,\ldots,N\}$ such that $|I\ra$ is the state with spin $\uparrow$ at positions $i_1,\ldots,i_k$ and spin $\downarrow$ everywhere else. We can then introduce the wavefunctions of off-shell Bethe states in this basis,
\be
S_I(\sigma_1,\ldots,\sigma_k)  \propto \la I | \sigma_1,\ldots,\sigma_k\ra \, ,
\label{eq:intro-stable}
\ee
which provide a set of generators for the twisted chiral ring. Nekrasov has proposed a physical definition of the corresponding twisted chiral ring elements as `orbifold defects'~\cite{Nekrasovtalk}. In this paper, we explain how to implement this orbifold construction in the $A$-twisted supersymmetric gauge theory to compute correlation functions of the twisted chiral operators~\eqref{eq:intro-stable} . We furthermore demonstrate that these operators are orthonormal with respect to the $A$-model two-point functions, corresponding to the fact that $\la I | J \ra = \delta_{I,J}$ in the spin chain Hilbert space.

\begin{figure}[htp]
\centering
\includegraphics[height=3.25cm]{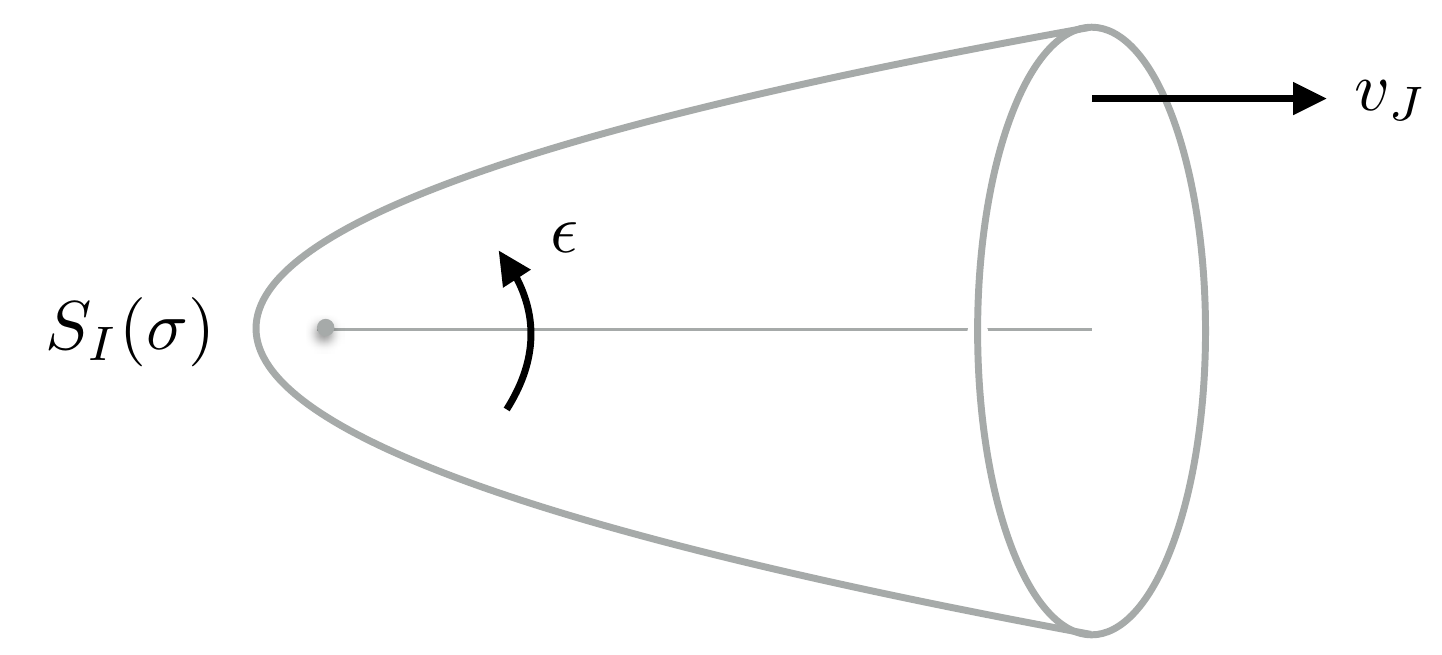}
\caption{Wavefunctions of spin chain eigenstates from $A$-twisted cigar partition functions.}
\label{fig:intro-cigar}
\end{figure}

In the algebraic Bethe ansatz, the eigenstates of the spin chain Hamiltonian are obtained by evaluating the off-shell Bethe state $|\sigma_1,\ldots,\sigma_k\ra$ on a solution of the Bethe equations~\eqref{eq:Bethe.eqn}. The functions $S_I(\sigma_1,\ldots,\sigma_k)$ evaluated on solutions of the Bethe equations are therefore the wavefunctions of the eigenstates in the `up-down' basis $|I\ra$. We will show that this wavefunction can be obtained directly from the supersymmetric gauge theory by computing the $A$-model in a cigar geometry with a vacuum corresponding to $I$ at infinity. More precisely, we first introduce an $\Omega$-background and then compute a normalized correlation function that is finite in the limit $\ep \to 0 $. This is shown in figure~\ref{fig:intro-cigar}.

\begin{figure}[htp]
\centering
\includegraphics[height=3.5cm]{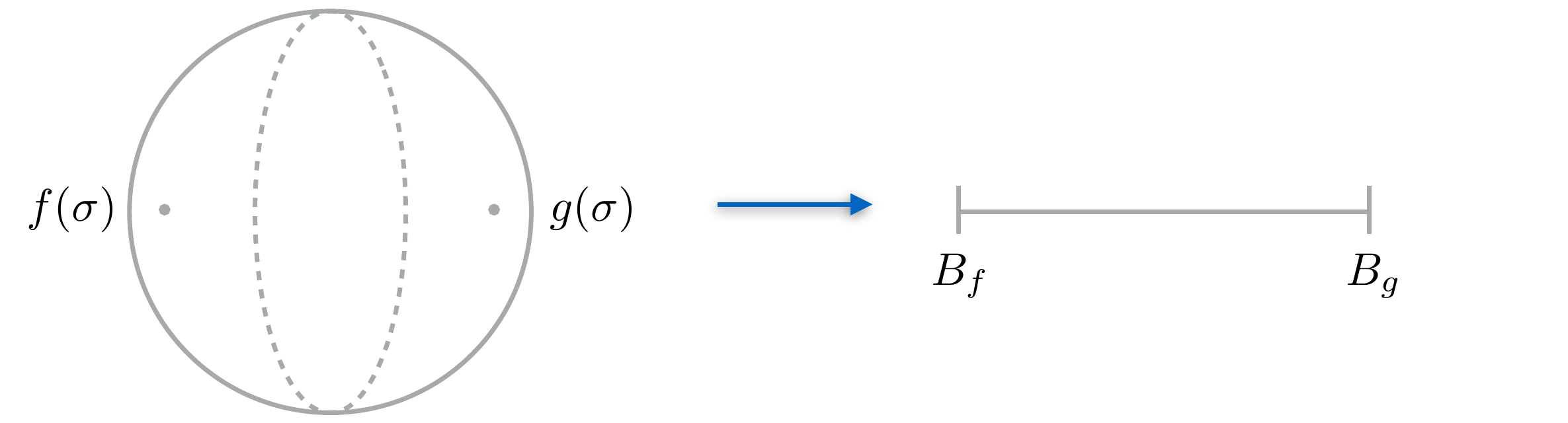}
\caption{Supersymmetric quantum mechanics formulation of $A$-model correlation functions in the limit $q\to 0$.}
\label{fig:qmlimit-intro}
\end{figure}

The parameter $q$ determines the quasi-periodic boundary condition for the spin chain. Apart from the final step of evaluating the off-shell Bethe states on solutions of the Bethe equations, the steps in the algebraic Bethe ansatz are independent of the parameter $q$. It is therefore sufficient to understand these aspects in the limit $q \to 0$, which corresponds to discarding instanton corrections in the $A$-twisted supersymmetric gauge theory. In this limit, correlation functions can be understood in a finite-dimensional $\cN=4$ supersymmetric quantum mechanics with boundary conditions preserving the same pair of supercharges as the $A$-twist. In particular, each twisted chiral operator generates a boundary condition in the supersymmetric quantum mechanics, and two-point functions  are computed by interval partition functions - as shown in figure~\ref{fig:qmlimit-intro}. In particular, we will provide two independent constructions of the boundary conditions generated by the operators $S_I(\sigma_1,\ldots,\sigma_k)$, either by coupling Neumann boundary conditions to additional degrees of freedom or as `thimble' boundary conditions.

The setup described above is compatible with turning on background holonomies for flavour symmetries. In the supersymmetric quantum mechanics description, background flavour holonomies around the circle become `real mass parameters' for flavour symmetries. The ordering of the holonomy eigenvalues or real masses can be identified with an ordering of sites on the spin chain. It is therefore natural to consider `Janus' interfaces which permute the ordering of the masses. We will show that $A$-model correlation functions of such Janus interfaces in between the elements $S_I(\sigma_1,\ldots,\sigma_k)$ reproduce matrix elements of the spin chain R-matrix. The Yang-Baxter relation is interpreted as the statement that a given permutation of real mass parameters can be decomposed in a number of ways into elementary Janus interfaces permuting a pair of real mass parameters. 

Finally, the Bethe/Gauge correspondence provides a physical realization of a parallel developments in geometry and representation theory, and many of the objects we consider here have already appeared in this context. The starting point is the statement that the twisted chiral ring is the equivariant quantum cohomology ring of the vacuum manifold of the supersymmetric gauge theory, $T^*G(k,N)$. The A-model correlation functions considered here can be formulated in the language of quasi-maps to the vacuum manifold. In particular, the functions $S_I(\sigma_1,\ldots,\sigma_k)$ were introduced in work of Maulik and Okounkov~\cite{Maulik:2012wi} as the `stable basis' in the quantum equivariant cohomology (see~\cite{Aganagic:2017gsx} for connections to Bethe wavefunctions). This paper is largely motivated by understanding these mathematical constructions in the language of supersymmetric gauge theory.

The paper is organized as follows. In Section~\ref{sec:primer} we collect some relevant properties of the Heisenberg spin chain. In Section~\ref{sec:setup} we describe 2d $\cN=(2,2)$ supersymmetric gauge theory and explain how to calculate their A-twisted sphere and cigar partition functions using supersymmetric localization. Section~\ref{sec:defects} focuses on the definition of the distinguished twisted chiral operators and their correlation functions. In Section~\ref{sec:qm} we reduce the problem to a 1d quantum mechanics and explain how the twisted chiral operators can arise from appropriate boundary conditions in this quantum mechanics. Finally, in Section~\ref{sec:r-matrix} we show how the spin chain R-matrices can be obtained as correlation functions of twisted chiral operators in the supersymmetric gauge theory. Our conventions and more technical details of the calculation are postponed to the Appendix.

\section{Spin Chain Primer}
\label{sec:primer}

In this section we collect some basic information on the Heisenberg XXX$_{\frac{1}{2}}$ spin chain, where all spins transform in the fundamental representation of $\mathfrak{su}(2)$. Many of the statements we present here and in subsequent sections have a natural generalization to spin chains with higher representations, as well as to higher rank algebras. Our notation is designed to match that of supersymmetric gauge theory and therefore differs from standard integrability conventions.

\subsection{Heisenberg Spin Chain}

In order to define the Heisenberg spin chain we need to specify a Hamiltonian and a Hilbert space on which it acts. The Hilbert space of the spin chain is the $N$-fold tensor product of the fundamental representation of $\mathfrak{su}(2)$,
\begin{equation}
\label{Hilbert.space}
\mathcal{V}=\underbrace{\mathbb{C}^2 \otimes \mathbb{C}^2 \otimes  \ldots \otimes \mathbb{C}^2}_{\scriptstyle N-\mbox{\small times}}\,.
\end{equation} 
We introduce standard basis elements $|\uparrow\, \rangle$ and $|\downarrow\, \rangle$ for each spin chain site $\C^2$. There is then a natural basis for $\cV$ that is labelled by subsets $I = \{I_1,\ldots,I_k\} \subset \{1,\ldots,N\}$ such that
\be
| I\rangle=|\downarrow \ldots \underbrace{\uparrow}_{I_1}\ldots \underbrace{\uparrow}_{I_k}\ldots \downarrow\rangle \, .
\ee
We define an inner product by demanding that the basis vectors at each site are orthonormal
 \begin{equation}
  \langle\, \downarrow|\downarrow\,\rangle= \langle\, \uparrow|\uparrow\,\rangle=1, \qquad \langle\, \downarrow|\uparrow\,\rangle= \langle\, \uparrow|\downarrow\,\rangle=0\,,
 \end{equation}
 and naturally extending this definition to $\cV$. Any operator $A:\mathcal{V}\to \mathcal{V}$ can be represented a $2^N \times 2^N$ matrix of its expectation values between tensor products of $|\uparrow\, \rangle$ and $|\downarrow\, \rangle$.
 
 The (twisted, homogeneous) Heisenberg spin chain is defined by the Hamiltonian,
 \begin{equation}\label{Hamiltonian}
 H=\frac{1}{\hbar}\sum_{i=1}^N(\mathbb{I}_{i,i+1}-\,\mathbb{P}_{i,i+1})\,,
 \end{equation} 
with the twisted boundary condition,
\begin{equation}\label{boundary.conditions}
|\uparrow\rangle_{N+1}=|\uparrow \rangle_1\,,\qquad |\downarrow\rangle_{N+1}= q\ |\downarrow\rangle_1\,.
\end{equation} 
Later on, we will also introduce inhomogeneities for each spin chain site.

The Hamiltonian commutes with the operator counting up spins and therefore the Hilbert space can be decomposed into a direct sum of spaces with fixed number of excitations,
\begin{equation}
\mathcal{V}=\bigoplus_{k=1}^N \mathcal{W}_k\,.
\end{equation} 
The spectrum of the Heisenberg spin chain can be then found using the celebrated Bethe ansatz. In particular, the eigenvalues of the Hamiltonian $H$ are obtained from the dispersion relation 
\begin{equation}
E=\sum_{a=1}^k \frac{\hbar}{\frac{\hbar^2}{4}-\sigma_a^2}\,,
\end{equation}
where we sum over the rapidities\footnote{It is common to use the letter $u$ to denote rapidities. In this paper we use the letter $\sigma$ instead in order to make connection with the gauge theory side of our story in the following sections.} of excitations $\sigma_a$, $a=1,\ldots,k$, which are solutions to the Bethe equations
\begin{equation}\label{Bethe-equations}
\left(\frac{\sigma_a+\frac{\hbar}{2}}{\sigma_a-\frac{\hbar}{2}} \right)^N= q\,\prod_{b\neq a}\frac{\sigma_a-\sigma_b+\hbar}{\sigma_a-\sigma_b-\hbar}\,, \qquad a=1,\ldots ,k\,.
\end{equation}

There is a natural generalisation of the homogeneous spin chain described above to include inhomogeneities $m_i$ at each site of the spin chain\footnote{It is common to use the letters $v_i$ to denote inhomogeneities. In this paper we use the letters $m_i$ instead in order to make connection with the gauge theory side of our story in the following sections.}. In that case we denote the spin chain Hilbert space as
\begin{equation}\label{Hilbert.space.inhom}
\mathcal{V}_m=\mathbb{C}^2_{m_1} \otimes \mathbb{C}^2_{m_2} \otimes  \ldots \otimes \mathbb{C}^2_{m_N}\,,
\end{equation} 
and the Bethe equations turn into
\begin{equation}\label{eq:Bethe.eqn2}
\prod_{i=1}^N\frac{\sigma_a-m_i+\frac{\hbar}{2}}{\sigma_a-m_i-\frac{\hbar}{2}} = q\,\prod_{b\neq a}\frac{\sigma_a-\sigma_b+\hbar}{\sigma_a-\sigma_b-\hbar}\,, \qquad a=1,\ldots ,k.
\end{equation}
For a given number of excitations $k$, there are $\binom{N}{k}$ solutions distinct solutions of the Bethe equations. The solutions $\sigma_a^I$ can be labelled by a subset $I=\{I_1,\ldots,I_k\}\subset\{1,\ldots,N\}$ such that, expanding around $q \to 0$, the solutions are of the form $\sigma_a^I=m_{I_a}-\tfrac{\hbar}{2}+\cO(q)$.

\subsection{R-matrices}
\begin{figure}[ht!]
\centering
\begin{subfigure}{.49\textwidth}
\centering
\includegraphics[height=3cm]{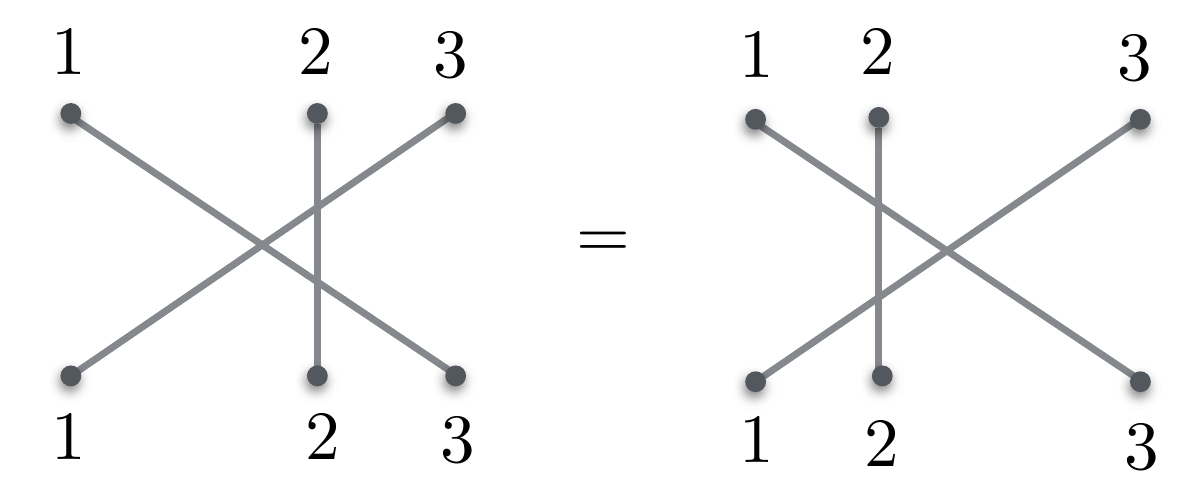}
\caption{Yang-Baxter equation.}
\label{Fig:YangBaxter.equation}
\end{subfigure}
\begin{subfigure}{.49\textwidth}
\centering
\includegraphics[height=3cm]{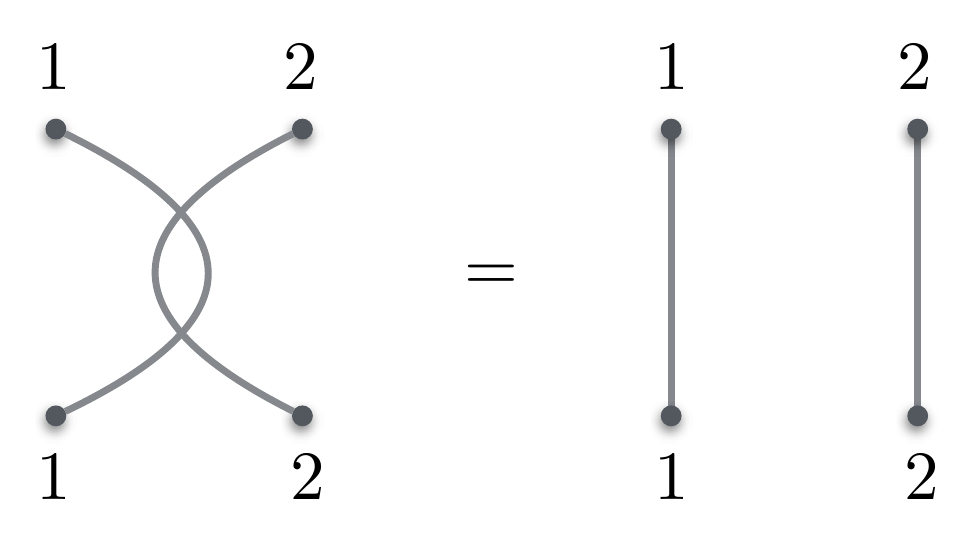}
\caption{Unitarity of the R-matrix.}
\label{Fig:unitarity}
\end{subfigure}
\caption{Relations satisfied by the R-matrix.}
\end{figure}

There are many independent ways to arrive at the Bethe equations \eqref{eq:Bethe.eqn2}. Usually, the most powerful method is the algebraic Bethe ansatz, which is based on the construction of an R-matrix. For the inhomogeneous spin chain, this is an operator acting on two sites,
\begin{equation}
R_{ij}(m_j-m_i):\mathbb{C}^2_{m_i} \otimes \mathbb{C}^2_{m_j} \to \mathbb{C}^2_{m_j} \otimes \mathbb{C}^2_{m_i}\,.
\end{equation}
It has rational dependence on $m_j-m_i$ and satisfies the regularity property $R_{ij}(0)\sim \mathbb{P}_{ij}$ where $\mathbb{P}_{ij}$ is the permutation operator, together with the Yang-Baxter equation (shown graphically in figure~\ref{Fig:YangBaxter.equation})
\begin{equation}\label{Yang.Baxter}
R_{12}(m_2-m_1)R_{13}(m_3-m_1)R_{23}(m_3-m_2)=R_{23}(m_3-m_2)R_{13}(m_3-m_1)R_{12}(m_2-m_1)\,,
\end{equation}
where $R_{ij}$ acts non-trivially only on $\C^2_{m_i}$ and $\C^2_{m_j}$.

An explicit form of the R-matrix is
\begin{equation}\label{Rmatrix.explicit}
R_{ij}(m_{ji})=\frac{1}{m_{ji}+\hbar}\left(m_{ji}\, \mathbb{I}_{ij}+\hbar \,\mathbb{P}_{ij}\right)\,,
\end{equation}
where $m_{ji}=m_j-m_i$ and we fixed the normalisation by the unitarity condition (shown graphically in figure~\ref{Fig:unitarity})
\begin{equation}\label{unitarity}
R_{12}(m)R_{21}(-m)=\mathbb{I}\,.
\end{equation}

\subsection{Bethe States}
\label{sec:bethe-states}

We now introduce an auxiliary space $\mathbb{C}_\sigma^2$ with spectral parameter $\sigma$ and define the monodromy matrix
 \begin{equation}
 M(\sigma)=R_{10}(\sigma-m_1)R_{20}(\sigma-m_2)\ldots R_{N0}(\sigma-m_N)\,,
 \end{equation}
 where each $R_{i0}(\sigma-m_i)$ acts non-trivially only on the auxiliary space and $\mathbb{C}^2_{m_i}$. 
 This is a $2\times 2$ matrix in the auxiliary space
 \begin{equation}
 M(\sigma)=\left( \begin{array}{cc}A(\sigma)&B(\sigma)\\C(\sigma)&D(\sigma)\end{array}\right)\,,
 \end{equation}
 where each matrix element is an operator acting on the Hilbert space \eqref{Hilbert.space.inhom}. Namely, each matrix element of $M(\sigma)$ in the auxiliary space can itself be represented by a $2^N\times 2^N$ matrix.
 
 For a given $k$, we define an off-shell Bethe state by
 \begin{equation}\label{eq.offshell}
 | \sigma_1 ,\ldots ,\sigma_k\rangle =B(\sigma_1)\ldots B(\sigma_k)|\Omega\rangle\,,
 \end{equation}
where $|\Omega \rangle=|\downarrow \ldots \downarrow\rangle$. Additionally, for given subset $I=\{I_1,\ldots,I_k\}$ of $\{ 1,\ldots , N\}$, we define the functions $S_{I}(\sigma)$ as the overlaps of off-shell Bethe state $|\sigma_1,\ldots,\sigma_k\rangle$ with the basis vectors:
\begin{equation}\label{stable.SC}
S_{I}(\sigma)=(-1)^{|I|} \cN(\sigma) \langle I| \sigma_1,\ldots,\sigma_k\rangle\,.
\end{equation}
Here we have introduced a normalization factor
\be
\cN(\sigma) = (-1)^{\tfrac{k(k-1)}{2}+k N}\frac{\prod_{a,i}(\sigma_a-m_i+\frac{\hbar}{2})}{\prod_{a,b}(\sigma_a-\sigma_b+\hbar)}\,,
\label{eq:normfactor}
\ee
which is independent of $I$. These functions can be computed explicitly with the result,
\begin{equation}\label{offshell.functions}
S_{I}(\sigma)=\mbox{Sym}_{\sigma}\frac{\prod\limits_{a=1}^{k}\left( \prod\limits_{i=1}^{I_a-1}(\sigma_a-m_{i}+\frac{\hbar}{2})\prod\limits_{i=I_a+1}^{N}(-\sigma_a+m_{i}+\frac{\hbar}{2})\right)}{\prod\limits_{a<b}(\sigma_a-\sigma_b)(\sigma_a-\sigma_b-\hbar)}\,.
\end{equation}
It can be shown that the states $|\sigma_1,\ldots,\sigma_k\ra$ become eigenstates of the spin chain Hamiltonian provided $\sigma_a$ are evaluated on a solution $\sigma_a^J$ of the Bethe equations. The functions $S_I(\sigma^J)$ are then (up to normalization) the wavefunctions of the Bethe eigenstates in the position basis $|J\ra$. In the following sections, we will explain how to construct such wavefunctions in the Bethe/Gauge correspondence.


\section{Setup}
\label{sec:setup}

In this section, we review the computation of correlation functions in $A$-twisted supersymmetric gauge theories on $\mathbb{CP}^1$ and a cigar. We review two approaches to computing such correlation functions using supersymmetric localization. The first leads to a contour integral in the complex Cartan subalgebra of the gauge group. The second is via equivariant localization on the moduli space of quasi-maps into the vacuum manifold. This will provide a foundation for the results presented in the following sections.

\subsection{The Model}
\label{sec:model}

\begin{figure}[ht!]
\centering
\includegraphics[height=1.8cm]{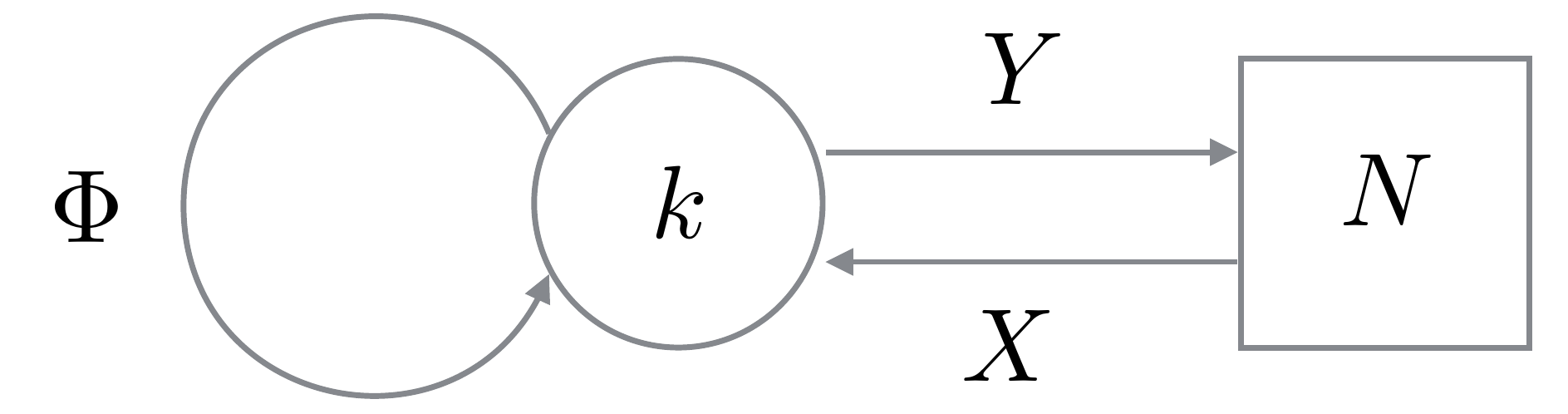}
\caption{Quiver}
\label{fig:quiver}
\end{figure}

We consider 2d $\cN=(2,2)$ supersymmetric gauge theories with R-symmetry \mbox{$U(1)_V \times U(1)_A$} that flow to sigma models onto cotangent bundles to complex Grassmannians, $T^*G(k,N)$. Such a theory has gauge symmetry $G = U(k)$ and flavour symmetry \mbox{$G_f = PSU(N) \times U(1)_\hbar$}. The field content is depicted in figure~\ref{fig:quiver} and can be summarized as follows:
\begin{itemize}
\item A vectormultiplet containing bosonic fields $(A_\mu,\sigma,D)$ transforming in the adjoint representation of $U(k)$, where $A_\mu$ is the gauge field, $\sigma$ is a complex scalar and $D$ is an auxiliary scalar.
\item Chiral multiplets $(\Phi,X,Y)$ transforming as shown in the table below.
\begin{table}[h]
\begin{center}
\begin{tabular}{c|c|c|c|c}
& $U(k)$ & $U(1)_V$ & $PSU(N)$ & $U(1)_\hbar$ \\ \hline
$\Phi$ & $\mathrm{adj}$  & $2$ & $1$ & $-1$ \\
$X$ & $\square$ & $0$ & $\overline\square$ & $ +\frac{1}{2}$ \\
$Y$ & $\overline\square$  & $0$ & $\square$ & $+\frac{1}{2}$
\end{tabular}
\end{center}
\end{table}
\item A superpotential $W = \tr(\Phi X Y)$ of the required $U(1)_V$ R-charge $+2$.
\item A complex twisted chiral parameter $t = \frac{\theta}{2\pi}+ i r$, combining a real FI parameter $r>0$ and theta angle $\theta$.
\end{itemize}
Our conventions are summarized in Appendix~\ref{app:conv}.

For positive FI parameter, $r>0$, the theory flows to a sigma model onto the moduli space of solutions the vacuum equations
\begin{gather}
\label{eq:D-term}
 \mu_\R - r \, \mathbb{1}  = 0 \,,\qquad [\sigma , \bar\sigma] = 0 \,,\\
\label{eq:F-term}
\Phi \cdot X = 0 \,,\qquad Y \cdot \Phi = 0 \,,\qquad X\cdot Y = 0 \,, \\
\sigma \cdot X = 0\,, \qquad - Y \cdot \sigma = 0\,, \qquad [\sigma,\Phi] = 0 \,,
\label{eq:vector}
\end{gather}
modulo constant gauge transformations. We define
\be
\mu_{\R} := X \cdot X^\dagger - Y^\dagger \cdot Y  + [ \Phi, \Phi^\dagger ]\,,
\ee
to be the moment map for the gauge symmetry. It can be shown that solutions require $\Phi=0$ and $\sigma = 0$, and that the remaining equations reproduce the hyper-K\"ahler quotient construction of $T^*G(k,N)$ where $r$ is the K\"ahler parameter of the base Grassmannian $G(k,N)$. We refer to this as the vacuum manifold $\vacman$.

It is useful to provide an algebraic description of the vacuum manifold. For $r>0$, we can replace the D-term equation~\eqref{eq:D-term} by the stability condition that the matrix $X$ has maximal rank and divide by complex gauge transformations,
\be
\vacman = \{ X , Y | X \cdot Y = 0 , \mathrm{rk}(X) = k \} / GL(k,\C) = T^*G(k,N) \, .
\ee
From this perspective, $X$ defines a $k$-plane in $\mathbb{C}^N$ corresponding to a point in the base Grassmannian $G(k,N)$. For example, in the case $k=1$, we have $T^*\mathbb{CP}^{N-1}$ with homogeneous coordinates $[X_1,\ldots,X_N]$ on the base. For negative FI parameter $r <0$, the roles of $X$ and $Y$ would be interchanged.

We can also introduce complex mass parameters for the flavour symmetry $G_f$
by coupling to a background vectormultiplet and introducing non-zero vacuum expectation values $(m_1,\ldots,m_N,\hbar)$ to the complex scalar in the vectormultiplet in a Cartan subalgebra $T_f \subset G_f$. Note that we must have $\sum_jm_j = 0$. In the presence of complex masses, equations~\eqref{eq:vector} are replaced by
\be
(\sigma_a -m_j+\tfrac{\hbar}{2} ) X^a{}_j = 0\,, \qquad (- \sigma_a -m_j+\tfrac{\hbar}{2} ) Y^j{}_a = 0\,, \qquad
(\sigma_a - \sigma_b +\hbar)\Phi^a{}_b = 0\,,
\ee
where $a=1,\ldots,k$ and $j=1,\ldots,N$ are gauge and flavour indices respectively  and $(\sigma_1,\ldots,\sigma_k)$ denote the eigenvalues of the vectormultiplet scalar $\sigma$. For generic values of the complex masses $(m_1,\ldots,m_N,\hbar)$, the vacuum manifold $\vacman$ is lifted, leaving behind $ \binom{N}{k} $ isolated massive vacua
\be
v_I \qquad : \qquad \sigma_a = m_{I_a} - \frac{\hbar}{2}\,, \qquad X^a{}_j = \sqrt{r} \delta_{j,I_a} \,,\qquad Y^j{}_a = 0\,, \qquad \Phi^a{}_b = 0 \, ,
\ee
labelled by subsets $I = \{I_1,\ldots,I_k\} \subset \{ 1,\ldots,N\} $ of size $|I| = k$. The massive vacua can be identified with the fixed points of the infinitesimal $T_f$ action on the vacuum manifold $ \vacman = T^*G(k,N)$ generated by $(m_1,\ldots, m_N,\hbar)$ and correspond geometrically to the coordinate hyperplanes in the base $G(k,N)$.

\subsection{Sphere Partition Function}
\label{sec:omega-sphere}

\begin{figure}[ht!]
\centering
\includegraphics[height=3cm]{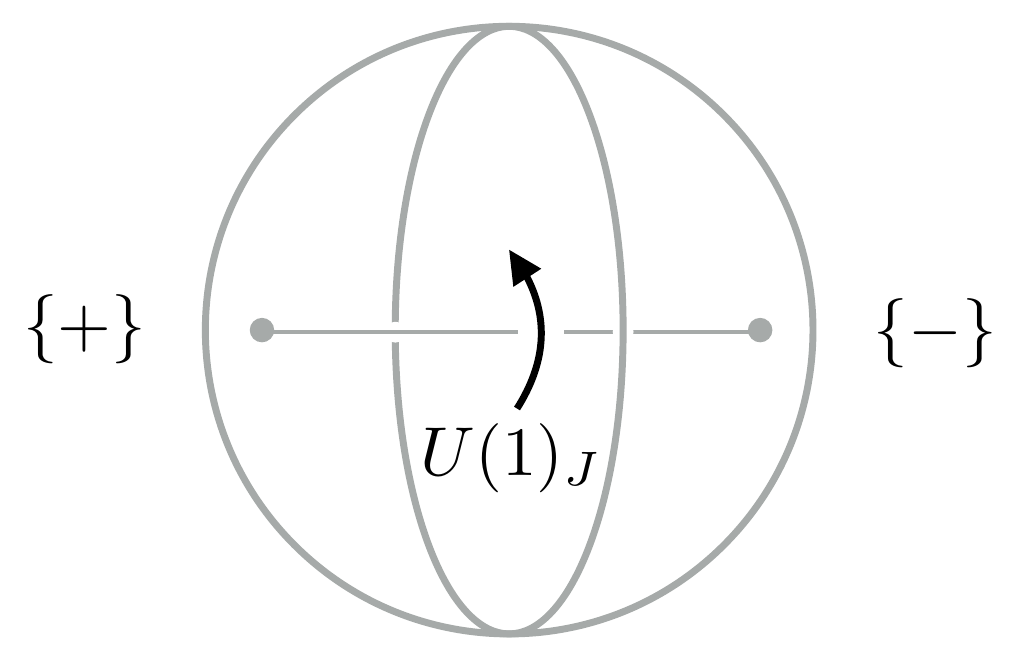}
\caption{We consider a $U(1)_J$ isometry of $\mathbb{CP}^1$ with fixed points $\{+\}$ and $\{-\}$. }
\label{fig:sphere}
\end{figure}

We will now consider correlation functions in the $\Omega$-deformed $A$-model on $C = \mathbb{CP}^1$, introduced in~\cite{Benini:2015noa,Closset:2015rna}. We introduce homogeneous coordinates $[z\!:\!w]$ on $\mathbb{CP}^1$ and define a $U(1)_J$ isometry that transforms the homogeneous coordinates by $(z,w) \to (e^{\ep/2}z,e^{-\ep/2}w)$ with fixed points $\{+\} = \{z=0\}$ and $\{-\} = \{w=0\}$, as shown in figure~\ref{fig:sphere}. The background preserves a pair of supercharges $\Q_-$, $\overline \Q_+$ that commute with the combination $U(1)_\ep := U(1)_J + U(1)_V$.

\subsubsection{Contour Integral}

Partition functions in the $\Omega$-deformed $A$-model can be computed exactly using supersymmetric localization for the supercharge $\Q = \Q_{-}+\overline \Q_+$~\cite{Benini:2015noa,Closset:2015rna}. This reduces the path integral to a contour integral over the complex Cartan subalgebra of $G$ parametrized by the eigenvalues $(\sigma_1,\ldots,\sigma_N)$ of the vectormultiplet scalar $\sigma$.

In order to express the contributions to the integrand of the contour integral from the various multiplets, it is convenient to introduce the following function
\bea
Z^{(n)}(\sigma) & = \frac{\Gamma_1(\sigma - \tfrac{n}{2}\ep )}{\Gamma_1(\sigma + \tfrac{n}{2}\ep+\ep)} 
 = \begin{cases}
\prod\limits_{\ell=-\tfrac{|n|}{2}+1}^{\tfrac{|n|}{2}-1}(\sigma+\ell \ep) \,,& \mathrm{if} \quad n <0 \,,\\
\prod\limits_{\ell=-\tfrac{n}{2}}^{\tfrac{n}{2}}(\sigma+\ell \ep)^{-1}\,, & \mathrm{if} \quad n \geq 0\,,
\end{cases}
\eea
where
\be
\Gamma_1(x) = \frac{ \ep^{x / \ep }}{\sqrt{2\pi \ep} } \Gamma(x/\ep)\,,
\ee
is Barnes' gamma function. Due to the functional equation $\Gamma_1(x+\ep) = x \,\Gamma_1(x)$, this ratio of Barnes' gamma functions is in fact a rational function of $\sigma$. 

The contribution to the integrand from a chiral multiplet of charge $r$ under the $U(1)_V$ vector R-symmetry and charge $q_f$ under a $U(1)_f$ flavour symmetry is $Z^{(q_fn-r)}(q_f\sigma)$, where $\sigma$ is the vectormultiplet scalar and $n\in\mathbb{Z}$ is the quantized flux through $\mathbb{CP}^1$.

Coming back to the model introduced in section~\ref{sec:model}, partition functions are expressed as a contour integral over the complex vectormultiplet scalar $\vec \sigma = (\sigma_1,\ldots,\sigma_k)$ together with a summation over the flux $\vec n = (n_1,\ldots,n_k) \in \mathbb{Z}^k$. The contributions to the integrand from the chiral multiplets are 
\bea
Z^{(\vec n)}_\Phi(\vec \sigma)  & =  \prod_{a,b=1}^k Z^{(n_{ab}-2)} (\sigma_{ab}-\hbar)\,,\\
Z^{(\vec n)}_X(\vec\sigma)  & = \prod_{i=1}^N\prod_{a=1}^k Z^{(n_a)}(\sigma_a-m_i+\tfrac{\hbar}{2})\,, \\
Z^{(\vec n)}_Y(\vec\sigma) & = \prod_{i=1}^N\prod_{a=1}^k Z^{(-n_a)}(-\sigma_a+m_i+\tfrac{\hbar}{2})\,,
\label{eq:1loop-chiral}
\eea
where we introduce a shorthand notation $\sigma_{ab}=\sigma_a - \sigma_b$ and $n_{ab} = n_a - n_b$. There is an additional contribution from the vectormultiplet
\bea
Z^{(\vec n)}_V(\vec\sigma) & = \prod_{a \neq b} Z^{(-n_{ab}-2)} (\sigma_{ab}) \\
& = \prod_{a < b}(-1)^{n_{ab}+1} \left(  (\sigma_a - \sigma_b)^2 - \frac{\ep^2}{4} (n_a-n_b)^2 \right)\,.
\label{eq:1loop-vector}
\eea
The partition function is then given by
\be
\la 1\ra_{S^2} = \sum_{n \in \mathbb{Z}^k } q^{\sum_a n_a}\int_\gamma \frac{d^k\sigma}{k!} \, \cZ^{(\vec n)}_{N,k}(\vec\sigma)\,,
\label{eq:sphere-pf}
\ee
where 
\begin{equation}
\cZ^{(\vec n)}_{N,k}(\vec\sigma)=(-1)^P Z^{(\vec n)}_V(\vec\sigma) Z^{(\vec n)}_\Phi(\vec\sigma) Z^{(\vec n)}_X(\vec\sigma)Z^{(\vec n)}_Y(\vec\sigma)\,,
\end{equation}
 combines the contributions from the vector and chiral multiplets. We include an additional sign $(-1)^P$ with $P = k^2 + (k+N+1)\sum_an_a$, where the factor $k^2$ fixes a sign ambiguity in the contributions from the chiral multiplets~\cite{Closset:2015rna} and $(k+N+1)\sum_an_a $ is an additional sign that can be absorbed into the definition of $q$. The contour $\gamma$ is given by the Jeffrey-Kirwan prescription, which reduces for $r>0$ to the contour surrounding poles at
\be
\sigma_a = m_i -\frac{\hbar}{2} - \left( \frac{n_a}{2} - \ell \right) \ep\,,
\ee
for all $i=1,\ldots,N$ and $\ell=0,\ldots,n_a$ coming from the contributions from the chiral multiplets $X^a{}_i$ in the fundamental representation of the gauge group. The summation over fluxes can therefore be restricted to the region $\vec n \in \mathbb{Z}^k_{\geq 0}$. We will often use the shorthand notation $n := \sum_a n_a$.

The partition function is enriched by inserting twisted chiral operators annihilated by $\Q_-$ and $\overline \Q_+$ at $\{\pm\}$. We will consider gauge-invariant functions $f(\vec \sigma)$ of the vectormultiplet scalar. As explained in~\cite{Benini:2015noa,Closset:2015rna}, there are then additional contributions
\bea
\{+\} \qquad & : \qquad f(\vec\sigma - \tfrac{\vec n}{2}\ep) \,,\\
\{-\} \qquad & : \qquad f(\vec \sigma + \tfrac{\vec n}{2}\ep)\,,
\eea
to the integrand in equation~\eqref{eq:sphere-pf}. We denote a correlation function with $f(\vec\sigma)$ inserted at $\{+\}$ and $g(\vec\sigma)$ inserted at $\{-\}$ by \begin{equation}\label{correlation.function}
\la f(\vec\sigma),g(\vec\sigma) \ra_{S^2}=\sum_{\vec n \in \mathbb{Z}^k_{\geq0} } q^{\sum_i n_i}\int_\gamma \frac{d^k\sigma}{k!} \, \cZ^{(\vec n)}_{N,k}(\vec\sigma)f(\vec\sigma - \tfrac{\vec n}{2}\ep)g(\vec \sigma + \tfrac{\vec n}{2}\ep)\,.
\end{equation}

Importantly, the contributions from $n_a>0$ vanish unless certain conditions are satisfied. For example, in the abelian case instanton corrections to $\la f(\sigma) , g(\sigma) \ra_{S^2}$ vanish unless the combined degree of the polynomials is greater than or equal to $2N-1$. This follows from the fact that for $n>0$ the only potential pole outside of the contour is at $\sigma \to \infty$, which only exists if $\deg(f)+\deg(g) \geq 2N-1$. This phenomenon can be understood as the condition to cancel the $U(1)_A$ axial anomaly. In all cases, the partition function $\la 1\ra_{S^2}$ receives contributions only from $\vec n=0$ and is therefore independent of $q$.

Moreover, correlation functions involving particular combinations of twisted chiral operators vanish, reflecting the structure of the twisted chiral ring. For example, in the abelian case
\bea
\la  f(\sigma) \prod_{j=1}^N(\sigma-m_j+\tfrac{\hbar}{2}) - q \,  f(\sigma-\ep)\prod_{j=1}^N(\sigma-m_j-\tfrac{\hbar}{2}) , g(\sigma) \ra_{S^2}  = 0\,, \\
\la f(\sigma)  , g(\sigma) \prod_{j=1}^N(\sigma-m_j+\tfrac{\hbar}{2}) - q \,  g(\sigma+\ep)\prod_{j=1}^N(\sigma-m_j-\tfrac{\hbar}{2})  \ra_{S^2} = 0 \,,
\eea
for any $f(\sigma)$ and $g(\sigma)$. In the limit $\ep \to0$, we recover the twisted chiral ring relations,
\be
\prod_{j=1}^N(\sigma-m_j+\tfrac{\hbar}{2}) - q \, \prod_{j=1}^N(\sigma-m_j-\tfrac{\hbar}{2}) = 0\,,
\ee
which coincide with the equivariant quantum cohomology ring of the vacuum manifold $\vacman = T^*\mathbb{CP}^{N-1}$.

In the limit $\ep\to0$, the general twisted chiral ring relations of a non-abelian theory that hold inside correlation functions are
\begin{equation}
\prod_{j=1}^N \frac{(\sigma_a-m_j+\frac{\hbar}{2})}{(\sigma_a-m_j-\frac{\hbar}{2})} =q\,\prod_{b\neq a}\frac{\sigma_a-\sigma_b+\hbar}{\sigma_a-\sigma_b-\hbar}\,, \qquad a=1,\ldots ,k.
\end{equation}
This coincides with the quantum equivariant cohomology ring of $\vacman = T^*G(k,N)$ and the Bethe equations \eqref{eq:Bethe.eqn2} for an inhomogeneous XXX$_{\frac{1}{2}}$ spin chain of length $N$ with quasi-periodic boundary conditions.

\subsubsection{Counting Quasi-maps}
\label{sec:quasi-map-sphere}

We now consider alternative approach to computing correlation functions in $A$-twisted gauged linear sigma models introduced in~\cite{Witten:1993yc,Morrison:1994fr} in terms of  vortex counting. This approach was derived rigorously from supersymmetric localization in~\cite{Closset:2015rna}. The mathematical formulation of this construction involves equivariant integrals over holomorphic `quasi-maps' to the vacuum manifold $\vacman$, which may be computed by equivariant localization~\cite{Okounkov:2015aa,Kim:2011aa}. This provides the link with recent mathematical work on the Bethe/Gauge correspondence~\cite{Aganagic:2017gsx}.

In this approach, we first set the complex mass parameters $(m_1,\ldots, m_N,\hbar)$ and the $\Omega$-deformation $\ep$ to zero, and consider configurations preserving both $\Q_-$ and $\overline \Q_+$. Such configurations are given by
\begin{gather}
   \mu_\R - r \mathbb{1} = - \frac{2 i}{g^2} F_{z\bar z} \label{eq:BPS} \,,\\
\overline D_{\bar z} X = 0\,, \qquad \overline D_{\bar z} Y = 0 \,,\qquad D_{\bar z} \Phi = 0\,,  \label{eq:BPS-2}\\
\Phi \cdot X = 0 \,,\qquad Y \cdot \Phi = 0 \,,\qquad X\cdot Y = 0 \,,
\label{eq:BPS-3}
 \end{gather}
together with
\begin{gather}
\sigma \cdot X = 0\,, \qquad - Y \cdot \sigma = 0 \,,\qquad [\sigma,\Phi] = 0 \,,
\label{eq:sigma-BPS-1}\\
D_z\sigma = 0\,,\qquad D_{\bar z} \sigma = 0\,, \qquad [\sigma,\bar\sigma] = 0\,,
\label{eq:sigma-BPS-2}
\end{gather}
modulo gauge transformations. 

The solutions of such `generalized vortex equations' are known as freckled instantons~\cite{Losev:1999tu,Losev:1999nt}. The moduli space of solutions has an algebraic description by dropping the D-term equation~\eqref{eq:BPS} in favour of a stability condition and dividing by complex gauge transformations.  This leads to a description in terms of stable `quasi-maps' from $C = \mathbb{CP}^1$ into the vacuum moduli space $\vacman = T^*G(k,N)$. The moduli space of solutions decomposes into a union of components
\be
\cM = \bigcup_{n \in \mathbb{Z}} \cM_n\,,
\ee
labelled by the vortex number or flux $n \in \mathbb{Z}$ through $\mathbb{CP}^1$, which coincides with the degree of the quasi-map. 

Now turning on the mass parameters $(m_1,\ldots, m_N,\hbar)$ and the $\Omega$-deformation parameter $\epsilon$ deforms the equations~\eqref{eq:sigma-BPS-1}-\eqref{eq:sigma-BPS-2} that determine $\sigma$ by replacing 
\be
\sigma \to \sigma + m + \hbar + \ep\,\cL_V\,,
\ee
where $(m_1,\ldots, m_N,\hbar)$ are understood to mean the infinitesimal $T_f$ flavour transformation generated by these parameters and $\cL_V$ is the Lie derivative along the vector field $V$ generating $U(1)_\ep$ rotations. This restricts the system to the fixed points of the corresponding $T_f$ action on the moduli space $\cM$. 

This can be understood as working equivariantly with respect to the action of $T_f \times U(1)_\ep$ on the moduli space $\cM$ with equivariant differential $ \Q = \Q_- + \overline \Q_+$. In particular, localization of the path integral to Gaussian fluctuations around $\Q_-$, $\bar\Q_+$-invariant configurations is equivalent to computing the following sum of equivariant integrals
\be
\la 1 \ra_{S^2} =\sum_{n\in \mathbb{Z}} q^n \int_{  [ \cM_n]^{\mathrm{vir}} } \mathbb{1}\,,
\ee
where $ [ \cM_n]^{\mathrm{vir}} $ is the virtual fundamental class. The correlation functions $\la f(\vec \sigma) , g(\vec\sigma) \ra_{S^2}$ correspond to computing the equivariant integrals of certain virtual equivariant cohomology classes $[f]$ and $[g]$ on $\cM_n$.

We will first explain how to compute the partition function in this manner in the abelian case, before considering the general case.

\subsubsection*{Abelian Case}

We first set the mass parameters $(m_1,\ldots, m_N,\hbar)$ and $\Omega$-deformation $\ep$ to vanish. Assuming $r >0$, we then have $\sigma = \Phi = 0$ and the remaining equations become
\begin{gather}
  \sum_{j=1}^N ( |X|^2 - |Y_j|^2 )  - r=- \frac{2 i}{g^2} F_{z\bar z}  
\label{eq:ab-D}\,, \\
\overline D_{\bar z} X_j = 0 \,,\qquad \overline D_{\bar z} Y_j = 0 \,,\qquad \sum_{j=1}^N X_jY_j = 0 \, .
\label{eq:ab-F}
\end{gather}
Solutions are labelled by the flux
\be
n = \frac{1}{2\pi} \int_C F \in \mathbb{Z}\,,
\ee
through $C = \mathbb{CP}^1$ and we denote the corresponding moduli space by $\cM_n$.

It is convenient to introduce the following algebraic description of the moduli space $\cM_n$. We first remove the $D$-term equation~\eqref{eq:ab-D} and replace it for $r>0$ by the stability condition that $X_j \neq 0$ for all $j = 1,\ldots, N$ except at a finite number of points on $\mathbb{CP}^1$. In addition, we divide by complex gauge transformations that leave the remaining equations~\eqref{eq:ab-F} invariant. A point in $\cM_n$ is now specified  by $N$ holomorphic sections $(X_j,Y_j)$ of $\cO(n) \oplus \cO(-n)$, such that $\sum_jX_j Y _j= 0$ and the sections $X_j$ are not all zero. It is now straightforward to compute the moduli spaces explicitly:
\begin{itemize}
\item If $n<0$, the moduli space is empty $\cM_n = \emptyset$.  
\item If $n=0$, we recover the algebraic description of the vacuum manifold $\cM_0 = \vacman = T^*\mathbb{CP}^{N-1}$. 
\item If $n>0$, we have $Y_j=0$ and the moduli space is parametrized by $N$ holomorphic sections $X_j$ of $\cO(n)$. Using a complex gauge transformation to set $A_{\bar z}= 0$, the holomorphic sections are homogeneous polynomials
\be
X_j(z,w) = \sum_{r=0}^n x_{j,r} z^{n-r}w^r \, .
\label{eq:Xexpansion}
\ee
The moduli space is therefore parametrized by the $N(n+1)$ coordinates $x_{j,r}$ that are not all zero, modulo residual constant $\mathbb{C}^*$ gauge transformations preserving $A_{\bar z} = 0$. We therefore find that $\cM_n = \mathbb{CP}^{N(n+1)-1}$. 
\end{itemize}

We now consider the fluctuations around a point on the moduli space $\cM_n$ for $n\geq0$. On general grounds such fluctuations decompose into chiral and Fermi multiplets with respect to the supersymmetry algebra generated by $\Q_-$ and $\bar\Q_+$. A 2d $\cN=(2,2)$ chiral multiplet of $U(1)_V$ charge $r$ transforming as a section of a line bundle $L$ contributes:
\begin{enumerate}
\item Chiral multiplets: $H^0(C,K_C^{r/2} \times L)$.
\item Fermi multiplets: $H^1(C,K_C^{r/2} \times L)$.
\end{enumerate}
Here, $K_C$ is the canonical bundle of the Riemann surface $C$. For us, $K_C = \cO(-2)$. This can be summarized by the statement that the fluctuations of a 2d $\cN=(2,2)$ chiral multiplet contribute $H^\bullet(C,K_C^{r/2} \times L)$ to the `virtual tangent bundle' of the moduli space.

Turning on $(m_1,\ldots, m_N,\hbar)$ and $\epsilon$ corresponds to working equivariantly with respect to the action of $T_f \times U(1)_\ep$ on $\cM_n$. Let us consider the fluctuations from each chiral multiplet in turn for $n> 0$, leaving the special case $n=0$ until the end.
\begin{itemize}
\item The fluctuations from each $X_j$ transform in $H^\bullet(C, \cO(n))$. There are therefore $N(n+1)$ chiral multiplets corresponding to fluctuations of the coordinates $x_{j,r}$ in equation~\eqref{eq:Xexpansion} and no Fermi multiplets. Under a $G \times G_f \times U(1)_\ep$ transformation generated by parameters $(\sigma,m_1,\ldots,m_N,\hbar,\ep)$, they transform with weight
\be
\sigma -m_j + \frac{\hbar}{2} + \left(\frac{n}{2}-s\right) \ep  \,,  \qquad j=1,\ldots, N\,,\qquad  s = 0,\ldots, n\,.
\ee
\item The fluctuations from each $Y_j$ transform in $H^\bullet(C, \cO(-n))$. There are therefore no chiral multiplets and $N(n-1)$ Fermi multiplets corresponding to fermion zero modes in the vortex background. They transform with weight
\be
-\sigma + m_j +\frac{\hbar}{2} - \left(\frac{n-2}{2} - s\right) \ep   \,,\qquad j=1,\ldots, N\,,\qquad  s = 0,\ldots, n-2\,.
\ee
\item The fluctuations from $\Phi$ transform in $H^\bullet(C,\cO(-2))$. There is therefore a single Fermi multiplet transforming with weight $-\hbar$.
\end{itemize}
In addition there is a contribution $-H^\bullet(\cO)$ from the vectormultiplet. Combining these contributions, the equivariant index of the virtual tangent bundle is
\be
T^{\mathrm{vir}}\cM_n= \sum_{j=1}^N \left[ \sum_{s =0}^{n} e^{\sigma -m_j + \frac{\hbar}{2} + (\frac{n}{2} - s) \ep} - \sum_{s = 0}^{n-2}  e^{-\sigma +m_j + \frac{\hbar}{2} - (\frac{n-2}{2}- s ) \ep} \right] - e^{-\hbar} - 1\,,
\ee
where $-1$ comes from the vectormultiplet.

The moduli space $\cM_n = \mathbb{CP}^{N(n+1)-1}$ has isolated fixed points under a generic $T_f \times U(1)_\ep$ transformation generated by $(m_1,\ldots,m_N,\hbar,\ep)$, which correspond to the $N(n+1)$ coordinate lines. We can label the fixed points by the pair $(i,r)$ with $i = 1\ldots, N$ and $r = 0,\ldots,n$. The fixed points correspond to sections
\be
X_{j}(z,w) = \delta_{ij} z^{n-r}w^r\,,
\ee
whose transformation under $(m_1,\ldots,m_N,\hbar,\ep)$ is compensated by a gauge transformation by $\sigma = \sigma|_{i,r} := m_i - \frac{\hbar}{2} - (\frac{n}{2} - r)\ep $. The equivariant index at the fixed point $(i,r)$,
\be
T^{\mathrm{vir}}_{(i,r)}\cM_n  =  \sum_{j=1}^N \left( \sum_{s=0}^n e^{m_i -m_j + (r-s) \ep} - \sum_{s = 1}^{n-1}  e^{-m_i+m_j+\hbar -(r-s)\ep} \right)- e^{-\hbar} -1 \,,
\ee
by the replacement $\sigma \to \sigma|_{i,r}$.

The contribution to the partition function from fluctuations around each fixed point of $\cM_n$ is encompassed in the virtual localization formula
\be
\int_{[\cM_n]^{\mathrm{vir}}} \mathbb{1} = \sum_{(i,r)} \frac{1}{e(T^\mathrm{vir}_{(i,r)}\cM_n)}\,,
\ee
where we have introduced the replacement rule $ e : \sum_i n_i e^{w_i} \to \prod_i w_i^{n_i} $ to compute the equivariant Euler character.
 This result is most neatly expressed as the following contour integral
\be
\int_\gamma d\sigma (-\hbar) \prod_{j=1}^N  \frac{\prod\limits_{s=1}^{n-2}\left(-\sigma +m_j + \frac{\hbar}{2} - \left(\frac{n-2}{2}- s \right) \ep\right)}{\prod\limits_{s=0}^{n} \left(\sigma -m_j + \frac{\hbar}{2} + \left(\frac{n}{2} - s\right) \ep\right)}\,,
\ee
where the contour surrounds the poles corresponding to the fixed points $\sigma = \sigma|_{i,r} = m_i - \frac{\hbar}{2} - (\frac{n}{2} - r)\ep $. This exactly reproduces the coefficient of $q^n$ for $n>0$ in the contour integral formula~\eqref{eq:sphere-pf}. Note that the Jeffrey-Kirwan residue corresponds to computing residues at poles of the integrand corresponding to fixed points of $\cM_n$. 

Let us now consider the special case $n=0$. The moduli space now corresponds to constant maps to the vacuum manifold $\vacman = T^*G(k,N)$ with bosonic fluctuations from both $X$ and $Y$. As above, the equivariant localization expression is neatly expressed as a contour integral
\be
\int_{[\cM_0]^{\mathrm{vir}}} \mathbb{1} =   \int_\gamma d\sigma(-\hbar)\prod_{j=1}^{N} \frac{1}{(\sigma-m_j+\tfrac{\hbar}{2} )(-\sigma+m_j+\tfrac{\hbar}{2} )} \, ,
\ee
where the contour surrounds the poles at $\sigma = m_j-\tfrac{\hbar}{2}$ from the contribution of $X_j$. This is a regular equivariant integral of $\mathbb{1}$ over the vacuum manifold $\vacman = T^*G(k,N)$.

The extension to include twisted chiral operators inserted at $\{\pm \}$ will be discussed in detail in section~\ref{Sec:stable.basis}.

\subsubsection*{Non-Abelian Case}

With gauge group $U(k)$, we again pass to an algebraic description of the moduli space $\cM$ of solutions to equations~\eqref{eq:BPS}-\eqref{eq:BPS-3} by removing the $D$-term equation in favour of a stability condition and dividing by complex gauge transformations. We therefore consider only
\begin{gather}
\overline D_{\bar z} X = 0 \,,\qquad \overline D_{\bar z} Y = 0\,, \qquad X\cdot Y = 0 \,,
\end{gather}
with the stability condition that the $k \times N$ matrix $X$ has maximal rank away from isolated points on $C = \mathbb{CP}^1$ and modulo complex $GL(k,\C)$ gauge transformations. A point in $\cM_n$ is then specified by:
\begin{itemize}
\item A holomorphic $GL(k,\C)$ bundle $V$.
\item Holomorphic sections $X$ and $Y$ of associated vector bundles $V \times \overline W$ and $\overline V \times W$ where $W \simeq C \times \mathbb{C}^N$ is a trivial vector bundle associated to the fundamental representation of the $PSU(N)$ flavour symmetry.
\item Constraints $X \cdot Y=0$.
\item Stability condition that $\mathrm{rk}(X) = k$ except at isolated points.
\end{itemize}

According to a theorem of Grothendieck, on $C = \mathbb{CP}^1$ we can decompose
\be
V = \cO(n_1) \oplus \cdots \oplus \cO(n_k) \qquad \sum_a n_a = n\,,
\ee
such that $X^a$ become sections of $\cO(n_a) \times \overline W$ and $Y_a$ become sections of $\cO(-n_a) \times W$. This leads to a stratification of the moduli space for flux $n  \in \mathbb{Z}$ into components labelled by integers $(n_1,\ldots,n_k) \in \mathbb{Z}^k$ with $\sum_a n_a = n$. The moduli space is empty if $n_a < 0$ for any $a=1,\ldots, k$. We therefore restrict attention to the region $n_a \in \mathbb{Z}_{\geq 0}$.

Fluctuations around a point on the moduli space $\cM_n$ decompose into chiral and Fermi multiplets with respect to the superalgebra generated by $\Q_-$ and $\bar\Q_+$. Following the discussion above, the contributions can be summarized by the equivariant index
\bea
T^{\mathrm{vir}} = H^\bullet(V \times \overline W \times W_{\hbar}^{\frac{1}{2}})  + H^\bullet(\overline V \times W \times W_{\hbar}^{\frac{1}{2}}) 
+ H^\bullet(K_C\times V \times \overline V \times W_{\hbar}^{-1}) - H^\bullet (V \times \overline V) \,,
\eea
where we have introduced yet another trivial line bundle $W_\hbar \simeq C \times \C$ associated to the fundamental representation of the flavour symmetry $U(1)_\hbar$. The first three contributions arise from the fluctuations of the chiral multiplets $X$, $Y$ and $\Phi$ respectively. The final contribution $-H^\bullet(V \times \bar V)$ is the contribution from the vectormultiplet.

The equivariant index is straightforward to write down explicitly for any $\vec n \in \mathbb{Z}^k_{\geq 0}$. Let us write the formula for the case when $n_a \geq 1$ for all $a = 1,\ldots,k$, and $n_a\neq n_b$ for $a\neq b$:
\bea
T^{\mathrm{vir}}\cM_n&= \sum_{j=1}^N \sum_{b=1}^k \left[ \sum_{s =0}^{n_b} e^{\sigma_b -m_j + \frac{\hbar}{2} + \left(\frac{n_b}{2} - s\right) \ep} - \sum_{s = 0}^{n_b-2}  e^{-\sigma_b +m_j + \frac{\hbar}{2} - \left(\frac{n_b}{2}- s-1 \right) \ep} \right] \\
&+ \sum_{b,c=1}^k \left[\sum_{s=0}^{n_{bc}-2} e^{\sigma_{bc}-\hbar +\left(\frac{n_{bc}}{2}-s-1\right)\ep }-\sum_{s=0}^{n_{bc}}e^{\sigma_{cb}-\hbar +\left(\frac{n_{bc}}{2}-s\right)\ep } \right]\\
&- \sum_{b,c=1}^k\left[ \sum_{s=0}^{n_{bc}} e^{\sigma_{bc} - \left(\frac{n_{bc}}{2} -s\right)\ep} -\sum_{s=0}^{n_{bc}-2}e^{\sigma_{cb} - \left(\frac{n_{bc}}{2} -s-1\right)\ep}  \right].
\eea

The moduli spaces themselves for $k>1$ are singular and do not admit an explicit description as in the abelian case. It is nevertheless possible to compute the equivariant fixed points in terms of the algebraic data and apply the virtual localization formula to compute the partition function. The fixed points are labelled by a decomposition $\vec n = \{n_1,\ldots,n_k\}$, a choice of vacuum $I = \{I_1,\ldots, I_k\} \subset \{1,\ldots,N\}$ and a vector $\vec s = \{ s_1,\ldots, s_k \}$ where $s_a \in \{ 0,1,\ldots, n_a\}$. The vectormultiplet scalar takes the following value at this point
\be
\sigma_a = \sigma_a|_{I,\vec s}  \equiv m_{I_a} - \frac{\hbar}{2} - \left(\frac{n_a}{2}-s_a\right)\ep\,,
\ee
and the virtual localization formula is
\be
\int_{[\cM_n]^{\mathrm{vir}}} \mathbb{1} = \sum_{|\vec n| = n} \sum_{(I,\vec s)} \frac{1}{e(T^\mathrm{vir}_{(\vec n,I,\vec s)}\cM_n)}\,.
\ee
This reproduces coefficient of $q^n$ for $n>0$ in the contour integral formula~\eqref{eq:sphere-pf} where the data $\{ \vec n , I , \vec s\}$ enumerate poles of the integrand chosen by the Jeffrey-Kirwan description. The case $n =0$ should again be treated separately and reproduces a regular equivariant integral over the vacuum manifold $\vacman = T^*G(k,N)$.

\subsection{Vortex Partition Function}

\begin{figure}[htp]
\centering
\includegraphics[height=3.25cm]{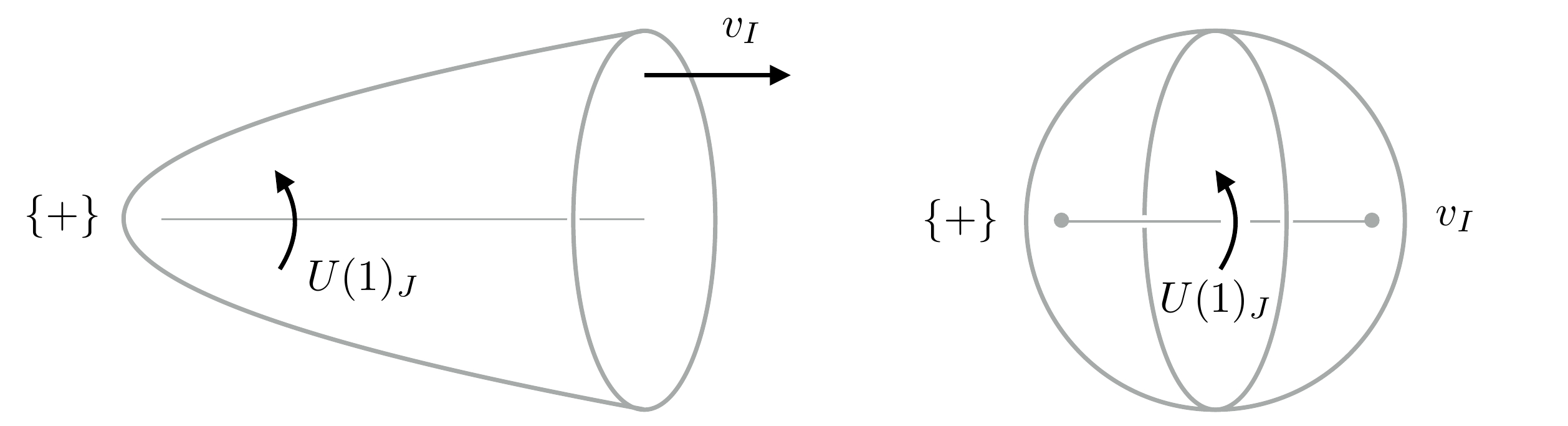}
\caption{We consider the vortex partition function on a cigar with $U(1)_J$ isometry and vacuum $v_I$ at infinity. This can also be viewed as a sphere with a fixed vacuum at $\{-\}$.}
\label{fig:cigar}
\end{figure}

We will also consider the vortex partition function or `cigar' partition function with a fixed vacuum $v_I$ at infinity. We can equivalently view this as a sphere with the boundary condition that the system sits in the vacuum $v_I$ at $\{ - \}$, as shown in figure~\ref{fig:cigar}. In section~\ref{sec:bethe-wav}, this partition function will be used to construct the wavefunctions of spin chain Bethe eigenstates. As above, we present the partition function both as a contour integral over the complex Cartan subalgebra of the gauge group and its interpretation in terms of counting quasi-maps that are `based' at $\{-\}$.

\subsubsection{Contour Integral}

Let us first consider the abelian case. The partition function with the vacuum $v_i$ at $\{-\}$ can be expressed as a contour integral in the vectormultiplet scalar $\sigma$,
\be
\la 1 \ra_{v_i}=\frac{1}{\Gamma_1(\hbar)} \int_{\gamma_i} d\sigma \, \tilde q^{-\frac{\sigma}{\ep} } \prod_{j=1}^N \Gamma_1(\sigma-m_j+\tfrac{\hbar}{2}) \Gamma_1(-\sigma+m_j+\tfrac{\hbar}{2}) \, .
\ee
where $\tilde q = (-1)^N q$. The integrand has poles at $\sigma = m_j - \tfrac{\hbar}{2} - \ell \ep$ for all $j=1,\ldots,N$ and $\ell \in \mathbb{Z}_{\geq 0}$. The contour $\gamma_i$ selects only those poles with $j = i$ arising from the 1-loop determinant for the chiral multiplet $X_i$ that has a non-zero expectation value in the vacuum $v_i$. 

The classical and 1-loop contributions can be factored out by normalizing by the value of the partition function at $q \to 0$
\be
\frac{\la 1 \ra_{v_i} }{ \la 1 \ra_{v_i} |_{q\to0} } =  \cV_i(q)\,.
\ee
The result,
\bea
 \qquad \cV_i(q) & = \sum_{ n =0}^\infty q^n   \prod_{j=1}^N \prod_{\ell=1}^n \frac{m_i-m_j-\hbar-(\ell-1)\ep}{m_i-m_j-\ell \ep}\,,
 \label{eq:vortex1}
\eea
is the vortex partition function with vacuum $v_i$ at infinity. 

This can be generalized to the non-abelian case with vacuum $v_I$,
\be
\la 1\ra_{v_I} = \int_{\gamma_I} \frac{d^k\sigma}{k!} \tilde q^{-\sum_a\sigma_a / \ep } \frac{ \prod\limits_{a \neq b}\Gamma_1(\sigma_{ab}+\ep)  }{\prod\limits_{a,b=1}^k\Gamma_1(-\sigma_{ab}+\hbar)} \prod_{a=1}^k\prod_{j=1}^n \Gamma_1(\sigma_a-m_j+\tfrac{\hbar}{2}) \Gamma_1(-\sigma_a+m_j+\tfrac{\hbar}{2}) \, , 
\ee
where $\tilde q = (-1)^{N} q$. The same integrand appears in the computation of the hemisphere partition function with the boundary condition supported on the whole of $\vacman = T^*G(k,N)$~\cite{Hori:2013ika,Honda:2013uca}. However, the contour $\gamma_I$ surrounds only the poles arising from the 1-loop determinant of the chiral multiplets $X_i$ for all $i \in I$. As above, we can extract the corresponding vortex partition function $\cV_I(q)$, which we will not write down explicitly.

We will denote the correlation function of a twisted chiral operator $f(\vec\sigma)$ at $\{+\}$ in the background with a supersymmetric vacuum $v_I$ at infinity $\{-\}$ by $\la f(\vec\sigma)\ra_{v_I}$. In the limit that we remove the $\Omega$-deformation, $\ep \to 0 $, such correlation functions have the common asymptotic behavior
\be
\la f(\vec\sigma)\ra_{v_I} \longrightarrow e^{-\frac{1}{\ep}\widetilde W(\vec\sigma^I(q))}+\ldots\,,
\ee
where $\widetilde W(\vec\sigma)$ is the effective twisted superpotential and $\sigma^I_a(q)=m_{I_a}-\tfrac{\hbar}{2}+\cO(q)$ is the particular solution of the Bethe equations \eqref{eq:Bethe.eqn2} associated to the fixed vacuum $v_I$ at infinity. Therefore, normalizing by the vortex partition function, we find that
\be\label{eq:expectation.value}
 \lim_{\ep \to 0} \frac{\la f(\vec\sigma)\ra_{v_I}}{\la 1\ra_{v_I}} = f(\vec\sigma_I(q))\,,
\ee
is independent of $\ep$ and it evaluates the twisted chiral operator at a particular solution of the Bethe equations corresponding to a fixed vacuum $v_I$. In section~\ref{sec:bethe-wav}, evaluation of this expectation value for a particular class of functions $f(\vec\sigma)$ will be used to construct the wavefunctions of spin chain Bethe eigenstates.

\subsubsection{Counting Quasi-maps}

The vortex partition function can also be expressed as an equivariant integral over the moduli space of vortices with vacuum $v_I$. The algebraic description of this moduli space is in terms of `based' quasi-maps to the vacuum manifold $\vacman$ such that the point $\{ -\}$ is mapped to a fixed vacuum $v_I$. Let us denote the moduli space of based quasi-maps by $\cM_{v_I}$ with components $\cM_{n,v_I}$ labelled by a flux $n \in \mathbb{Z}$. It is straightforward to relate equivariant integrals over the moduli spaces $\cM_n$ and $\cM_{n,v_I}$ as follows.

Let us first consider the abelian case with a vacuum $v_i$ at the fixed point $\{-\}$ and flux $n>0$. In the notation of equation~\eqref{eq:Xexpansion}, the vacuum condition fixes the coordinates $x_{j,0} = \delta_{ij}$. Here, a complex gauge transformation has been used to set the non-zero coefficient to $1$. Therefore, the moduli space is $\cM_{n,v_i} = \mathbb{C}^{Nn}$, parametrized by the remaining coordinates $x_{j,\ell}$ for $j=1,\ldots,N$ and $\ell = 1,\ldots,n$. This has a single equivariant fixed point at the origin, with compensating gauge transformation $\sigma = \sigma_{i,0} = m_i - \tfrac{\hbar}{2} - \tfrac{n}{2}\ep $. 

Clearly, the vacuum condition removes bosonic fluctuations corresponding to the coordinates $x_{j,0}$. On the other hand, there are now additional fermionic zero modes. The result can be summarized by adding a contribution $-T_{v_i} \vacman$ to the virtual tangent bundle. The result is that
\be
\cV_i(q) = w_i \sum_{n\geq 0} q^n \int_{\gamma_{i,n}} \cZ^{(n)}_{N,1}(\sigma)\,,
\ee
where $Z^{(n)}_{N,k}(\sigma)$ is the same integrand that appeared in the computation of the sphere partition function \eqref{eq:sphere-pf} and the contour $\gamma_{i,n}$ computes the residue at the pole $\sigma = \sigma_{i,0} = m_i - \tfrac{\hbar}{2} - \tfrac{n}{2}\ep $. Finally, 
\be
w_i = \prod_{j\neq i}(m_i-m_j)(\hbar - m_i + m_j)\,,
\ee
is the equivariant weight of the tangent space $T_{v_i} \vacman$. This reproduces the vortex partition function~\eqref{eq:vortex1}.

This formula can be extended to the non-abelian case,
\be
\cV_I(q) = w_I \sum_{\vec n \in \mathbb{Z}_{\geq 0}^k} q^{\sum_a n_a} \int_{\gamma_{I,\vec n}} \cZ^{(\vec n)}_{N,k}(\vec\sigma)\,,
\ee
where the contour $\gamma_{I,\vec n}$ surrounds the poles at $\sigma_a = \sigma_a|_{I,\vec 0}  = m_{I_a} - \frac{\hbar}{2} - \frac{n_a}{2}\ep$ and
\be
w_I = \prod_{i \in I}\prod_{j\notin I} (m_i - m_j)(\hbar - m_i + m_j)  \,,
\ee
is the equivariant weight of the tangent space $T_{v_I} \vacman$.

\subsection{Factorization}

Although not strictly necessary for this paper, we note that partition functions on the sphere can be decomposed into vortex partition functions,
\bea
\la 1 \ra_{S^2} 
 =  \sum_I \cV_I(q,\ep) \frac{1}{w_I} \cV_I(q,-\ep)\,.
\eea
Note that $\ep$ appears with opposite sign in each factor because the neighbourhoods of $\{+\}$ and $\{-\}$ look like $\Omega$-deformations with opposite orientation. A similar result holds for correlation functions of twisted chiral operators. A nice feature of this expressions, is that it manifests the fact that the sphere partition function reproduces an equivariant integral over the vacuum manifold $\vacman = T^*G(k,N)$ in the limit $q \to0$.

\begin{figure}[htp]
\centering
\includegraphics[height=3cm]{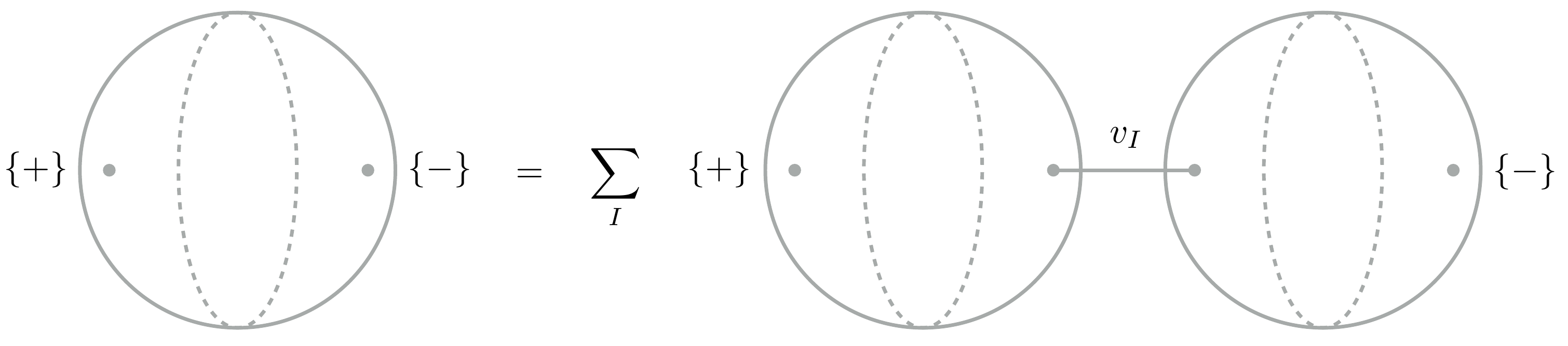}
\caption{The $\Omega$-deformed $A$-model partition function on a sphere can be decomposed as a sum over the vacua $v_I$ of products of vortex partition functions centered on $\{+\}$ and $\{-\}$.}
\label{fig:decomposition}
\end{figure}

Intuitively, the sphere is `pinched' to form a pair of spheres with identified marked points, as shown in figure~\ref{fig:decomposition}. The path integral is then decomposed into a product of integrals over moduli spaces of quasi-maps restricted to land on a fixed point $v \in \vacman$ at each marked point, which is then integrated over the vacuum manifold $\vacman$. In the presence of mass parameters $(m_1,\ldots,m_N,\hbar)$ and $\ep$, this becomes an equivariant integral over $\vacman$ and equivariant localization reproduces the above equation.

\section{Defect Operators in 2d}\label{Sec:stable.basis}
\label{sec:defects}

In this section, we consider in more detail the correlation functions of twisted chiral operators constructed from invariant functions of $\sigma$. In the presence of the mass parameters $(m_1,\ldots,m_N,\hbar)$, the twisted chiral ring coincides with the equivariant quantum cohomology of the vacuum manifold $\vacman = T^*G(k,N)$. In computing correlation functions, twisted chiral operators can be interpreted as equivariant cohomology classes on the moduli spaces $\cM_n$ of quasi-maps to $\vacman$.

We will study a distinguished set of generators for the twisted chiral ring, which coincide with the stable basis in equivariant cohomology introduced in~\cite{Maulik:2012wi} and coincide with the wavefunctions of off-shell Bethe states. In section~\ref{sec:twisted-chiral}, we focus on abelian theories, reviewing the dictionary between polynomials in $\sigma$ and equivariant cohomology classes, and introducing the stable basis elements in the case of $\vacman = T^*\mathbb{CP}^{N-1}$. In Section~\ref{sec:orbifold}, we consider a systematic physical construction of the stable generators using an orbifold construction proposed by Nekrasov~\cite{Nekrasovtalk} and explain how to implement this construction in the $A$-model to recover the functions $S_I(\vec\sigma)$. Finally, in section~\ref{sec:bethe-wav} we show that the correlation functions of the stable basis in a cigar background reproduce on-shell Bethe wavefunctions.

\subsection{Abelian Theories}
\label{sec:twisted-chiral}

In abelian theories, twisted chiral operators $f(\sigma)$ are polynomials in the vectormultiplet scalar $\sigma$. Turning on complex mass parameters, we restrict to polynomials that are homogeneous in the parameters $(\sigma, m_1,\ldots,m_N,\hbar)$. Recalling that these parameters have $U(1)_A$ charge $+2$, a homogeneous polynomial of degree $d$ will correspond to an equivariant form on $\vacman$ of degree $2d$.

Let us first consider the case of vanishing flux, where we have constant maps into the vacuum manifold, $\vacman = T^*\mathbb{CP}^{N-1}$. Consider the homogeneous polynomial 
\be
\sigma - m_j +\frac{\hbar}{2} \, ,
\ee
corresponding to the equivariant weight of the coordinate $X_j$. This is a polynomial representative of the equivariant cohomology class Poincar\'e dual to the submanifold $\{X_1 = 0\} \subset T^*\mathbb{CP}^{N-1}$. Similarly, the homogeneous polynomial
\be
\prod_{j=1}^r(\sigma-m_j+\tfrac{\hbar}{2})\,,
\ee
is a polynomial representative of the cohomology class Poincar\'e dual to the complex codimension $r$ submanifold $\{ X_1 = \cdots =X_r= 0 \}\subset T^*\mathbb{CP}^{N-1}$. Similar comments apply to polynomials that are products of weights of $X_j$'s and $Y_j$'s.

Let us study the special case $N=2$ more systematically. Recall that the vacuum manifold is found by solving the moment map constraints
\bea
\mu_\C  & = X_1Y_1 + X_2 Y_2= 0\,,  \\
\mu_\R  & = |X_1|^2+|X_2|^2 - |Y_1|^2 - |Y_2|^2 = r\,,
\eea
modulo $U(1)$ gauge transformations for $r >0$. The vacuum manifold is therefore $T^*\mathbb{CP}^1$ with the homogeneous coordinates $X_1,X_2$ on the base. The charges of these fields under the $U(1)$ gauge and $U(1)_m \times U(1)_\hbar$ flavour symmetries are shown in table~\ref{tab:charges}. We denote the mass parameter for $U(1)_m$ by $m = m_1 = - m_2$. 

\begin{table}[htp]
\centering
\begin{tabular}{c|c|c|c}
&$U(1)$ &$U(1)_{m}$&$U(1)_\hbar$\\
\hline
$X_1$&$1$&$-1$&$\tfrac{1}{2}$\\
$X_2$&$1$&$1$&$\tfrac{1}{2}$\\
$Y_1$&$-1$&$1$&$\frac{1}{2}$\\
$Y_2$&$-1$&$-1$&$\frac{1}{2}$
\end{tabular}
\caption{Gauge and flavour charges of chiral multiplets in the case $k=1$ and $N=2$.}
\label{tab:charges}
\end{table}

\begin{figure}[htp]
\centering
\includegraphics[height=2.75cm]{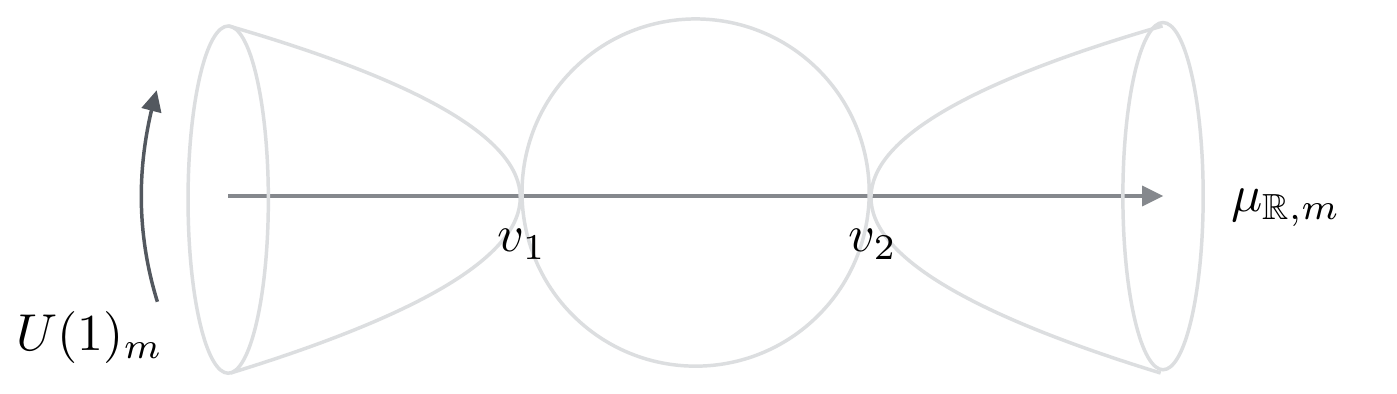}
\caption{The slice $\mu_{\C,m}=0$ of $T^*\mathbb{CP}^1$ exhibited as an $S^1$-fibration over $\mathbb{R}$ parametrized by $\mu_{\R,m}$. The $U(1)_m$ flavour symmetry rotates the fibers with fixed points at $\mu_{\R,m} = -r$ and $\mu_{\R,m} =r$, corresponding to the vacua $v_1$ and $v_2$ respectively.}
\label{fig:vac-sqed}
\end{figure}

It is convenient to exhibit the vacuum manifold as an $S^1$ fibration over $\mathbb{R}^3$ induced the the action of $U(1)_m$. The base is parametrized by the invariant real and complex moment maps for $U(1)_m$,
\bea
\mu_{\mathbb{C},m} & = -X_1 Y_1+X_2Y_2 \,,\\
\mu_{\mathbb{R},m} & = -|X_1|^2 + |X_2|^2 + |Y_1|^2 - |Y_2|^2 \, ,
\eea
while the fiber is parametrized by
\be
\vartheta = \frac{1}{2}\mathrm{arg}(X_1 / X_2) = - \frac{1}{2}\mathrm{arg}(Y_1 / Y_2)\,,
\ee
and rotated by $U(1)_m$. In figure~\ref{fig:vac-sqed}, we have drawn the slice $\mu_{\C,m}=0$ of the vacuum manifold. The fiber degenerates at the fixed points of the $U(1)_m$ action, $\mu_{\R,m} = -r$ and $\mu_{\R,m} = r$, corresponding to the positions of the supersymmetric massive vacua $v_1$ and $v_2$ respectively.

\begin{figure}[htp]
\centering
\includegraphics[height=6.5cm]{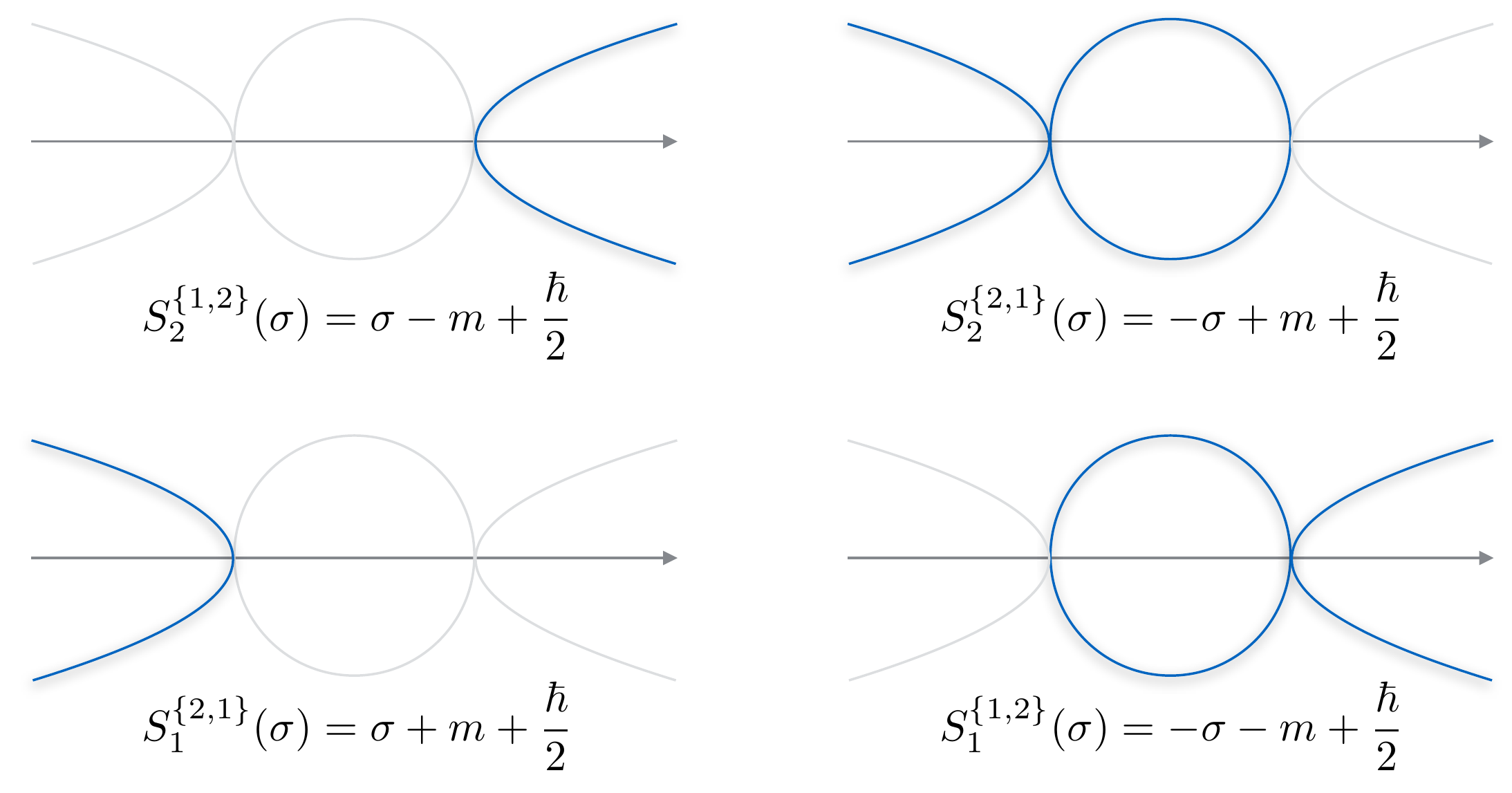}
\caption{The subspaces defined by setting one of the chiral multiplets $X_1$, $X_2$, $Y_1$, $Y_2$ to vanish and the polynomial representations of the Poincar\'e dual cohomology classes. For future reference, we have included the stable basis labels to be introduced in section~\ref{sec:orbifold}.}
\label{fig:sub-sqed}
\end{figure}

To each individual chiral multiplet there is a holomorphic lagrangian in $T^*\mathbb{CP}^1$ 
defined by the vanishing of the corresponding coordinate,
\begin{align}
&\{ X_1=0\}=F_2\,,\qquad
\{Y_1=0\}=\mathbb{CP}^1\cup F_1\,,\\
&\{ X_2=0\}=F_1\,,\qquad
\{Y_2=0\}=\mathbb{CP}^1\cup F_2\,,
\end{align}
where $F_1$, $F_2$ denotes the fibers of $T^*\mathbb{CP}^1$ at the points $v_1$,$v_2$ on the base $\mathbb{CP}^1$. These holomorphic lagrangian submanifolds are illustrated in figure~\ref{fig:sub-sqed}. The equivariant weights of the coordinates then provide polynomial representatives of the Poincar\'e dual cohomology classes. In particular, we have
\begin{align}
[F_2]=\sigma-m+\tfrac{\hbar}{2}\,,\qquad [\mathbb{CP}^1 \cup F_1]=-\sigma+m+\tfrac{\hbar}{2}\,,\\
[F_1]=\sigma+m+\tfrac{\hbar}{2}\,,\qquad [\mathbb{CP}^1 \cup F_2]=-\sigma-m+\tfrac{\hbar}{2}\,.
\end{align}

Certain pairs of these classes provide convenient bases for the equivariant cohomology of $T^*\mathbb{CP}^1$. Let us explain this statement by considering the pair
\bea
S_1(\sigma) & : = [\mathbb{CP}^1 \cup F_2] = - \sigma - m +\frac{\hbar}{2}\,,  \\
S_2(\sigma) & : = [F_2] = \sigma - m +\frac{\hbar}{2} \, .
\label{eq:N=2basis}
\eea
A general cohomology class is represented by a polynomial $f(\sigma)$ of degree at most $2$. This can be reduced to a polynomial of degree $1$ using the $q \to 0$ limit of twisted chiral ring relations,
\be
(\sigma - m +\tfrac{\hbar}{2})(\sigma + m +\tfrac{\hbar}{2}) = 0 \, .
\ee
It may then be expressed uniquely as
\be
f(\sigma) =  \frac{f(m-\tfrac{h}{2})}{-2m+h} S_1(\sigma) + \frac{f(-m+\tfrac{h}{2})}{-2m+h} S_2(\sigma)\,,
\ee
where the coefficients are rational functions of $m$ and $\hbar$. For example,
\be
[\mathbb{CP}^1] = S_1(\sigma)-S_2(\sigma) =  -2\sigma \, .
\ee
Note that we could have alternatively chosen the complementary pair of cohomology classes $[F_1]$ and $[\mathbb{CP}^1 \cup F_1]$ whose polynomial representatives are obtained from $S_1(\sigma)$ and $S_2(\sigma)$ by the replacements $m \to -m$ and $\sigma\to -\sigma$. 

The above discussion has a natural extension to $N>2$. The basis generalizing~\eqref{eq:N=2basis} is given by
\be
\label{eq:stable-basis-ab-2}
S_i(\sigma) = \prod_{j=1}^{i-1}(\sigma-m_j+\tfrac{\hbar}{2} )\prod_{j=i+1}^N(-\sigma+m_j+\tfrac{\hbar}{2} )\, ,
\qquad 1 = 1,\ldots, N \, ,
\ee
corresponding to the holomorphic lagrangians in $T^*\mathbb{CP}^{N-1}$ defined by 
\be
\begin{cases}
\, X_j  = 0 & \quad\mathrm{for}\quad j=1,\ldots,i-1\,, \\
\,\, Y_j = 0 & \quad\mathrm{for} \quad j=i+1,\ldots,N\,.
\end{cases}
\ee
These polynomials match the wavefunctions of the off-shell Bethe states \eqref{offshell.functions} for $k=1$. More generally, we will introduce a stable basis for each permutation $\pi$ of $\{1,\ldots,N\}$. They correspond to the holomorphic lagrangians obtained by setting to zero the coordinates $X_{\pi(j)}$ for $j=1,\ldots,i-1$ and $Y_{\pi(j)}$ for $j=i+1,\ldots,N$. The corresponding polynomial representatives $S^{(\pi)}_{i}(\sigma)$ are obtained by permuting the mass parameters $(m_1,\ldots,m_N)$ in expression~\eqref{eq:stable-basis-ab-2}. A more systematic approach, including the generalization this basis to the non-abelian case, is presented in section~\ref{sec:orbifold}.

Now let us consider correlation functions with flux $n>0$. Let us take a polynomial $f(\sigma)$ corresponding to an equivariant submanifold $Z \subset \vacman$. Inserting this operator at the point $\{+\}$ on $\mathbb{CP}^1$ leads to an additional contribution $f(\sigma - \tfrac{n}{2}\ep)$ to the integrand of the correlation function.  This corresponds to the equivariant cohomology class on the moduli space of quasi-maps, $\cM_n = \mathbb{CP}^{N(n+1)-1}$, that is Poincar\'e dual to the subspace of quasi-maps that land in $Z \subset \vacman$ at the point $\{+\}$. 

Let us demonstrate this in more detail. We recall that the moduli space $\cM_n = \mathbb{CP}^{N(n+1)-1}$ is parametrized by the coefficients $\{x_{j,\ell}\}$ of the sections
\be
X_j(z,w) = \sum_{\ell=0}^n x_{j,\ell} z^\ell w^{n-\ell}\,,
\ee
modulo complex rescalings. Upon restriction to the point $\{+\}$, we find
\be
X_j(0,w) = x_{j,0} w^n\,,
\label{eq:restrict}
\ee
where the coordinate $x_{j,0}$ has equivariant weight
\be
\sigma - m_j +\frac{\hbar}{2} - \frac{n}{2}\ep \, .
\ee
Inserting the twisted chiral operator $f(\sigma) = \sigma - m_j +\tfrac{\hbar}{2}$ at the point $\{+\}$ acts as an equivariant delta-function for this mode: it corresponds to the equivariant cohomology class Poincar\'e dual to $\{x_{j,0} = 0 \} \subset \cM_n$. From equation~\eqref{eq:restrict}, this is the subvariety corresponding to quasi-maps that land in $\{X_1 = 0\} \subset \vacman$ at $\{+\}$. Similar comments apply to $\{-\}$.

In addition to the bosonic fluctuations, there are fermionic fluctuations arising from the superpartners of $Y_j$. Inserting a polynomial corresponding to a weight of $Y_j$ corresponds to adding a fermionic zero mode localized at $\{+\}$. Such an insertion should be interpreted as a `virtual' cohomology class on the moduli space $\cM_n$.

\subsection{Orbifold Construction}
\label{sec:orbifold}

We now consider a more systematic definition of the stable basis of twisted chiral operators using an orbifold construction proposed by Nekrasov~\cite{Nekrasovtalk}. We will explain how to implement this construction in $A$-model correlation functions, reproducing the wavefunctions of off-shell Bethe states. Our approach mirrors similar computations for orbifold-type codimension-two defects in higher dimensions~\cite{Alday:2010vg,Kanno:2011fw,Bullimore:2014awa,Bullimore:2014upa}.

\subsubsection{Chiral Multiplet}

As a warm-up, we will first consider the case of a chiral multiplet. The contribution to the partition function of a chiral multiplet of $U(1)_V$ R-charge $r=0$ is
\bea
Z^{(q_fn)}(q_f\sigma) & = \frac{\Gamma_1(q_f(\sigma - \tfrac{n}{2}\ep ))}{\Gamma_1(q_f(\sigma + \tfrac{n}{2}\ep)+\ep)} \,,
\label{eq:chiral-pf}
\eea
where the chiral multiplet has charge $q_f$ under a $U(1)_f$ flavour symmetry with flux $n\in \mathbb{Z}$. In order to perform an orbifold construction independently at $\{+\}$ and $\{-\}$ we should first express this result in terms of contributions localized at these fixed points.

Let us show that the Barnes' gamma functions in the numerator and denominator arise from fluctuations in the neighbourhood of $\{+\}$ and $\{-\}$ respectively. Provided $n\geq0$, the only fluctuations on $\mathbb{CP}^1$ come from the complex boson $\phi$ in the chiral multiplet. A solution of $\bar D\phi = 0$ in a neighbourhood of $\{\pm\}$ takes the form 
\bea
\{+\} \quad : \quad \phi & = w^n \sum_{\ell \geq 0 } \phi_\ell^+(z/w)^\ell \,, \\
\{-\}  \quad : \quad \phi & = z^n \sum_{\ell \geq 0 } \phi_\ell^-(w/z)^\ell \, ,
\eea
whose coefficients have equivariant weights
\be
\phi_\ell^\pm  \quad : \quad \sigma|_{\pm} \pm \ell \ep \, ,
\ee
where $\sigma|_{\pm} = \sigma \mp \tfrac{n}{2}\ep$ is the value of the complex vectormultiplet scalar at $\{\pm\}$. It is convenient to encode these contributions in the equivariant indices
\bea
\cI_{+}  & = \sum_{\ell \geq 0 }e^{\sigma  -\frac{n}{2}\ep + \ell \ep }  = \frac{e^{q_f\sigma|_{+}}}{1-e^{\ep}} \, , \\
\cI_{-}  & = \sum_{\ell \geq 0 }e^{\sigma + \frac{n}{2} \ep - \ell \ep  }  = \frac{e^{q_f\sigma|_{-}}}{1-e^{-\ep}} \, .
\label{eq:index-chiral}
\eea
This manifests that the contributions from $\{+\}$ and $\{-\}$ are related by $\ep \to -\ep$, reflecting the fact that we have an $\Omega$-background with opposite orientation in the neighbourhoods of the fixed points $\{+\}$ and $\{-\}$. 

The contributions to the partition function are recovered from the equivariant index by taking expansion in positive powers of $e^{\ep}$ and replacing $\sum_i n_i e^{\omega_i} \to \prod_{i}\omega_i^{-n_i}$. In particular,
\bea
&\cI_+ && \longrightarrow \quad \Gamma_1(q_f(\sigma - \tfrac{n}{2}\ep) )  \,,\\
&\cI_- &&   \longrightarrow \quad \Gamma_1(q_f(\sigma + \tfrac{n}{2}\ep)+\ep)^{-1}\,,
\eea
where zeta-function regularization of infinite products $\Gamma_1(x) \sim \prod_{\ell \geq 0} (x+\ell\ep)^{-1}$
is understood. The total partition function~\eqref{eq:chiral-pf} combining the contributions from $\{+\}$ and $\{-\}$ is of course a rational function, reflecting the fact that there is a finite number of global holomorphic sections of $\cO(n)$ on $\mathbb{CP}^1$.

We now perform an orbifold construction at either of the fixed points $\{+\}$ or $\{-\}$ by choosing a subgroup $\mathbb{Z}_N \subset U(1)_J$ and turning on a discrete holonomy $g = \omega^{i-1}$ for the $U(1)_f$ flavour symmetry for some choice of $i=0,\ldots,N-1$ where $\omega^N = 1$. The introduction of this orbifold is implemented at the level of the equivariant index by replacing
\bea
\sigma|_{\pm} \to \sigma|_{\pm} + (i-1)\frac{\ep}{N} \quad , \quad \ep \to \frac{\ep}{N}  \,,
\eea
and then averaging over the transformation $\ep \to \ep + 2\pi i s$ for $s=0,\ldots,N-1$, keeping the expectation value $\sigma|_{\pm}$ fixed. Applying this operation to the equivariant indices~\eqref{eq:index-chiral} in the region $-N<q_f(i-1)<N$ we find that
\bea
\cI_{+} 
& \longrightarrow 
\begin{cases}
\cI_{+} - e^{q_f \sigma|_{+}} & \mathrm{if} \quad 0 < q_f(i-1) < N \,,\\
\cI_{+} & \mathrm{if} \quad -N < q_f(i-1) \leq 0\,,
\end{cases} \\
I_-  
& \longrightarrow
\begin{cases}
\cI_{-}  & \mathrm{if} \quad 0 \leq q_f(i-1) < N \,,\\
\cI_{-} - e^{q_f \sigma|_{-}} & \mathrm{if} \quad -N < q_f(i-1) < 0\,.
\end{cases}
\eea
Therefore, the partition function remains unchanged for $-N < q_f(i-1) \leq 0$ at $\{+\}$ and $0 \leq q_f(i-1) < N$ at $\{-\}$. Otherwise, the partition function is multiplied by an additional factor $q_f \sigma|_{\pm}$. In this case, the orbifold construction is then equivalent to inserting the twisted chiral operator $q_f\sigma$. 

Let us now explain this procedure by implementing the orbifold construction directly on the mode expansion. We focus on the fixed point $\{+\}$ and set the coordinate $w = 1$ with the understanding that $\phi$ transforms with weight $\sigma|_+ = \sigma - \frac{n}{2}\ep$ under $U(1)_f$ flavour transformations. This is natural because $\sigma|_+$ is the vacuum expectation value of the complex scalar at $\{+\}$. With this understanding, the chiral field is expanded
\be
\phi(z) = \sum_{\ell \geq 0} \phi_\ell^+z^\ell \, .
\ee

In the presence of the orbifold, the chiral field should transform under $\mathbb{Z}_N$ transformations as $\phi(z) \to g^{q_f}\phi(\omega z)$, where $g=\omega^{i-1}$. It is therefore convenient to define a deformed field $\tilde \phi(z) := z^{q_f(i-1)} \phi(z)$, which absorbs the effect of the discrete holonomy and transforms in the standard way $\tilde\phi(z) \to \tilde\phi(\omega z)$. This has an expansion
\be
\widetilde\phi(z) = \sum_{\ell \geq 0 } \phi_\ell^+ z^{q_f(i-1)+\ell} \, .
\ee
The modes that are invariant under the $\mathbb{Z}_N$ action are $\phi^+_\ell$ such that $q_f(i-1)+\ell = \tilde \ell N$ for some $\tilde\ell \in \mathbb{Z}$. Projecting onto $\mathbb{Z}_N$-invariant modes and redefining the complex coordinate $z\to z^{1/N}$, it is straightforward to see that
\be
\widetilde\phi(z) = \begin{cases}
\sum\limits_{\tilde \ell \geq 1} \widetilde \phi^+_{\tilde\ell} z^{\tilde \ell} & \mathrm{if} \quad 0 < q_f(i-1) < N\,, \\
\sum\limits_{\tilde \ell \geq 0} \widetilde \phi^+_{\tilde\ell} z^{\tilde \ell} & \mathrm{if} \quad -N < q_f(i-1) \leq 0\,,
\end{cases}
\ee
where we have defined $\widetilde \phi^+_{\tilde\ell}:=\phi^+_{\tilde\ell N-q_f(i-1)}$. Therefore the effect of the orbifold on the mode expansion is trivial if $-N < q_f(i-1) \leq 0$. However, in the region $0 < q_f(i-1) < N$ the orbifold has removed the mode $\phi^+_0$ and we should therefore multiply the integrand of the partition function by the corresponding equivariant weight $q_f\sigma|_+$.

\subsubsection{Gauge Theory}

We now consider the supersymmetric gauge theory introduced in section~\ref{sec:model} and introduce a $\mathbb{Z}_N$ orbifold with discrete holonomy that breaks the $U(k)$ gauge and $ PSU(N)$ gauge and flavour symmetry to a maximal torus~\cite{Nekrasovtalk}. The construction depends on the following data:
\begin{itemize}
\item A permutation $\pi$ of $\{1,\ldots,N\}$, which specifies the discrete holonomy for the flavour symmetry,
\begin{equation}
(g_F)^i{}_j = \omega^{\pi(i)-1}\delta^i{}_j\,.
\label{eq:flavour-holonomy}
\end{equation}
\item An ordered subset $I = \{I_1,\ldots,I_k\} \subset \{1,\ldots, N\}$ with $I_a < I_b$ for $a<b$, which specifies the discrete holonomy for the gauge symmetry,
\begin{equation}
(g_G)^a{}_b =  \omega^{I_{\pi(a)}-1}\delta^a{}_b \,.
\label{eq:gauge-holonomy}
\end{equation}
\end{itemize}
We first perform the orbifold construction at the point $\{+\}$. We will denote the twisted chiral operator introduced by this orbifold construction at $\{+\}$ by $S^{(\pi)}_{I}(\vec\sigma)$. In the following, in order to simplify our notation, we write formulae for the unit permutation $\pi=\{1,\ldots,N\}$ and define $S_{I}(\vec\sigma) := S^{\{1,\ldots,N\}}_{I}(\vec\sigma)$.
 
 The starting point for the computation is the equivariant index for contributions at $\{+\}$,
\be
\cI_{+} = \frac{1}{1-e^{\ep}}\left[ \sum_{i=1}^N \sum_{a=1}^k \left( e^{\sigma_a|_{+}-m_i+\tfrac{\hbar}{2}}+ e^{-\sigma_a|_{+}+m_i+\tfrac{\hbar}{2}} \right)  +\sum_{a,b=1}^k e^{\sigma_{ab}|_{+}-\hbar + \ep} - \sum_{a \neq b} e^{\sigma_{ab}|_{+} }  \right] \, .
\ee
The orbifold construction is implemented by shifting the complex flavour and gauge parameters according to the discrete holonomy,
\begin{equation}
	m_i \rightarrow m_i + (i-1)\frac{\epsilon}{N} \qquad \sigma_a|_{\pm} \rightarrow \sigma_a|_{\pm} + (I_a-1)\frac{\epsilon}{N}  \qquad   \ep \to \frac{\epsilon}{N} \ ,
\end{equation}
and averaging over the transformations $\ep \to \epsilon + 2\pi i s$ for $s=0,\ldots, N-1$. This operation leads to a modification of the equivariant index by 
\be
\cI_+ \to \cI_+ - \sum_{a=1}^k\left( \sum_{i=1}^{I_a-1} e^{\sigma_a|_+-m_i+\tfrac{\hbar}{2} } +\sum_{i=I_a+1}^{N} e^{-\sigma_a|_++m_i+\tfrac{\hbar}{2} }  -\sum_{a<b}\left(e^{\sigma_{ab}|_+} + e^{\sigma_{ab}|_+-\hbar}\right) \right) \, ,
\ee
and therefore to an insertion of 
\be
\label{eq:stable-basis}
S_I(\vec\sigma) = \mathrm{Sym}_\sigma \frac{\prod\limits_{a=1}^k\left(\prod\limits_{i=1}^{I_a-1}(\sigma_{a}-m_i+\tfrac{\hbar}{2} )\prod\limits_{i=I_a+1}^N(-\sigma_a+m_i+\tfrac{\hbar}{2} )\right)}{\prod\limits_{a<b}(\sigma_a-\sigma_b)(\sigma_a-\sigma_b-\hbar)} \,,
\ee
in the integrand of the partition function. In writing this expression, we have symmetrized over $\sigma_1,\ldots, \sigma_k$ as the operator is inserted inside a contour integral that is symmetric in these parameters. Note that in the abelian case $k=1$ with $I=\{i\}$, the above expression reduces to the abelian formula~\eqref{eq:stable-basis-ab-2} considered above.

\setlength\arrayrulewidth{0.7pt}
\begin{figure}[ht!]
\begin{center}\begin{tabular}{|p{0.2cm}|p{0.2cm}|p{0.2cm}|p{0.2cm}|p{0.2cm}|p{0.2cm}|p{0.2cm}|p{0.2cm}|p{0.2cm}|p{0.2cm}|}
\hline
\cellcolor{gray} &\raisebox{-0.1cm}{$\star$}&\cellcolor{lightgray}&\cellcolor{lightgray}&\cellcolor{lightgray}&\cellcolor{lightgray}&\cellcolor{lightgray}&\cellcolor{lightgray}&\cellcolor{lightgray}&\cellcolor{lightgray}\\[0.18cm]
\hline
\cellcolor{gray}&\cellcolor{gray}&\cellcolor{gray}&\raisebox{-0.1cm}{$\star$}&\cellcolor{lightgray}&\cellcolor{lightgray}&\cellcolor{lightgray}&\cellcolor{lightgray}&\cellcolor{lightgray}&\cellcolor{lightgray}\\[0.18cm]
\hline 
\cellcolor{gray}&\cellcolor{gray}&\cellcolor{gray}&\cellcolor{gray}&\cellcolor{gray}&\cellcolor{gray}&\raisebox{-0.1cm}{$\star$}&\cellcolor{lightgray}&\cellcolor{lightgray}&\cellcolor{lightgray}\\[0.18cm]
\hline
\cellcolor{gray}&\cellcolor{gray}&\cellcolor{gray}&\cellcolor{gray}&\cellcolor{gray}&\cellcolor{gray}&\cellcolor{gray}&\cellcolor{gray}&\raisebox{-0.1cm}{$\star$}&\cellcolor{lightgray}\\[0.18cm]
\hline
\end{tabular}
\end{center}
\caption{Table encoding the function $S_I(\sigma)$ for $N=10$, $k=4$ and $I=\{ 2,4,7,9\}$.}
\label{fig:table}
\end{figure}
This formula can be understood graphically as explained in figure~\ref{fig:table}. 
In order to read off the numerator of \eqref{eq:stable-basis} one draws a $k\times N$ table and indicates the positions of $\{I_1,\ldots,I_k\}$ by~$\star$. Then one fills in the cells to the right (light grey) with equivariant weights $-\sigma_{a}+m_i+\tfrac{\hbar}{2}$ corresponding to fields $Y^i{}_a$, and the cells to the left (dark grey) with weights \mbox{$\sigma_{a}-m_i+\tfrac{\hbar}{2}$} corresponding to fields $X^a{}_i$. Finally, one multiplies all the weights in the table. The denominator of \eqref{eq:stable-basis} is universal for all $I$ for a given $k$.

Performing the orbifold construction at $\{+\}$ with a holonomy labelled by a general permutation $\pi$ inserts the operator $S^{(\pi)}_{I}(\vec\sigma)$ given by the expression
\be\label{stable.basis.explicit}
S^{(\pi)}_{I}(\vec\sigma) = S_{\pi^{-1}(I)}(\vec\sigma) |_{m_j \mapsto m_{\pi(j)} } \, .
\ee
Finally, performing the same orbifold construction at the other fixed point $\{-\}$ leads to an insertion of the operator $S^{(\iota \cdot \pi)}_{I}(\vec\sigma)$ where $\iota : \{1,\ldots,N\} \to \{N,\ldots,1\}$ is the longest or reflection permutation.

We now perform the same computation at $\{+\}$ using a zero mode analysis, highlighting the additional features that appear compared with a single chiral multiplet. For simplicity, we restrict ourselves here to the zero flux sector. The computation in the general flux sector is obtained by replacing $\sigma_a \to \sigma_a|_+ = \sigma_a - \frac{n_a}{2}\ep$ everywhere with $\sigma|_+$ fixed under the orbifold action.

In absence of the orbifold defect, we can expand the fields $X$, $Y$ and $\Phi$ around the point $\{+\}$ as holomorphic functions of $z$,
\begin{equation}
	X^a{}_i(z) = \sum_{\ell=0}^\infty (x_\ell)^a{}_i \, z^\ell\,,   \qquad 	Y^i{}_a(z) = \sum_{\ell=0}^\infty (y_\ell)^i{}_a\, z^\ell  \,,\qquad \Phi^a{}_b(z) = \sum_{\ell=1}^\infty (\phi_\ell)^a{}_b\, z^\ell \, .
\end{equation}
Note that since the adjoint chiral multiplet $\Phi$ has $U(1)_V$ charge $+2$, the expansion of this field starts at $\cO(z)$ in the twisted theory.

In the presence of the $\mathbb{Z}_N$ orbifold with discrete holonomies~\eqref{eq:flavour-holonomy} and~\eqref{eq:gauge-holonomy} at $\{+\}$ we can introduce the deformed fields
\begin{eqnarray}
	\widetilde{X}^a_{  \ i}(z) &=& z^{I_a-i}\sum_{\ell=0}^\infty (x_\ell)^a_{\ i}\, z^\ell \ , \qquad 
	\widetilde{Y}^i_{ \ a}(z) = z^{i-I_a}\sum_{\ell=0}^\infty (x_\ell)^a_{\ i} \,z^\ell \ , \nonumber \\
	\widetilde{\Phi}^a_{\ b}(z) &=& z^{I_a - I_b}\sum_{\ell=1}^\infty(\phi_\ell)^a_{\ b} \, z^\ell \ ,
\end{eqnarray}
which transform under $\mathbb{Z}_N$ transformations by replacing their argument $z \to \omega z$. We now project onto modes in the expansion that are invariant under $\mathbb{Z}_N$. For example, the invariant modes in the expansion of $\widetilde{X}^a{}_i$ are parametrized by $(x_\ell)^a{}_{\ i} $ with $I_a-i +\ell = \tilde \ell N$ for some $\tilde\ell\in\mathbb{Z}$. Replacing $z \to z^{1/N}$ and relabelling the coefficients, the invariant parts of the expansions are
\begin{eqnarray}
	\widetilde{X}^a{}_{i}(z)&=& 
	\left\{ \begin{array}{ll}\sum\limits_{\ell=0}^\infty(\widetilde{x}_\ell)^a{}_{i}\, z^{\ell} & \ , \ I_a \le i \\
	\sum\limits_{\ell=1}^\infty(\widetilde{x}_\ell)^a_{ \ i}\, z^{\ell} & \ , \ I_a > i 
	\end{array}
	\right.
	\ \ , \qquad 
	\widetilde{Y}^i{}_{a}(z)= 
	\left\{ \begin{array}{ll}\sum\limits_{\ell=0}^\infty(\widetilde{y}_\ell)^a{}_{i}\, z^{\ell} & \ , \ I_a \ge i \\
	\sum\limits_{\ell=1}^\infty(\widetilde{y}_\ell)^a_{ \ i}\, z^{\ell} & \ , \ I_a < i 
	\end{array}
	\right. \ \ , \nonumber \\
	\widetilde{\Phi}^a{}_{b}(z) &=& 
	\left\{ \begin{array}{ll}\sum\limits_{\ell=0}^\infty(\widetilde{\phi}_\ell)^a_{ \ i}\, z^{\ell} & \ , \ a < b \\
	\sum\limits_{\ell=1}^\infty(\widetilde{\phi}_\ell)^a_{ \ i}\, z^{\ell} & \ , \ a \ge b 
	\end{array}
	\right. \ \ ,
\end{eqnarray}
near $\{+\}$.

The orbifold defect has completely changed the zero mode structure at $\{+\}$, as one can see from the above expansions. Note that before orbifolding, each component of $X$ and $Y$ had a zero mode at $\{+\}$, while $\Phi$ vanished there. However, the orbifold has eliminated some of the fluctuations of $X$ and $Y$ at $\{+\}$ with equivariant weights
\be
\prod\limits_{a=1}^k    \left(\prod\limits_{i=1}^{I_a-1}(\sigma_{a}-m_i+\tfrac{\hbar}{2} )\prod\limits_{i=I_a+1}^N(-\sigma_a+m_i+\tfrac{\hbar}{2} )\right) \, ,
\ee
while introducing additional zero modes for $\Phi$ with equivariant weights
\be
\prod\limits_{a<b}(\sigma_a-\sigma_b-\hbar)^{-1} \, .
\ee
Furthermore, the orbifold breaks the gauge symmetry $U(k) \rightarrow U(1)^k$ at $\{+\}$ and the broken generators develop additional zero modes in the defect background parametrizing the complete flag variety $\mathbb{F}_k = U(k)/U(1)^k$. This leads to an additional contribution
\be
\prod\limits_{a<b}(\sigma_a-\sigma_b)^{-1}\,,
\ee
corresponding to the equivariant weight at a fixed point of $\mathbb{F}_k$. This combines with the contribution from $\Phi$ to form the equivariant weight of the cotangent bundle $T^*\mathbb{F}_k$. The symmetrization over $\sigma_1,\ldots,\sigma_k$ together with these denominator factors can be interpreted as an equivariant integral over the moduli space $T^*\mathbb{F}_k$ of the defect. The appearance of a hyper-K\"ahler moduli space is expected since in the absence of the mass parameter $\hbar$ the defect preserves a $\cN=(0,4)$ supersymmetry. Combining these contributions reproduces the function $S_I(\sigma)$ obtained in equation~\eqref{eq:stable-basis}.

\subsection{Bethe Wavefunctions}
\label{sec:bethe-wav}

The functions $S_I(\sigma)$ are up to normalization the wavefunctions of the off-shell Bethe states for the spin chain in the up-down basis, $\la I | \sigma_1,\ldots,\sigma_k\ra$. The Bethe eigenstates themselves are obtained by evaluating the auxiliary variables on a solution $\sigma_a^J$ of the Bethe equations, $|\Psi_J\ra = | \sigma_1^J,\ldots,\sigma^J_k\ra$. The wavefunctions of the Bethe eigenstates in the up-down basis are then $\la I | \Psi_J\ra = (-1)^{|I|}\cN(\sigma^J)^{-1} S_I(\sigma^J)$ where the normalization factor $\cN(\sigma)$ is defined in equation~\eqref{eq:normfactor}.

These wavefunctions arise in the supersymmetric gauge theory from expectation value of the stable basis elements $S_I(\sigma)$ in the cigar background with vacuum $\sigma^J$ at infinity, as in equation~\eqref{eq:expectation.value}. More precisely, we compute the limit $\epsilon\to 0$ of a normalized correlation function with $S_I(\sigma)$ inserted at the tip of cigar and vacuum $v_J$ at infinity,
\be
\label{eq:onshell.Bethe.states}
\lim_{\ep \to 0} \frac{\la S_I(\vec\sigma)\ra_{v_J}}{\la 1\ra_{v_J}} =S_{I}(\sigma^J) \, .
\ee
We have already mentioned that this evaluates the function $S_I(\sigma)$ at the solution to Bethe equations \eqref{eq:Bethe.eqn2} associated to the vacuum $v_J$, with expansion $\sigma_a^J=m_{J_a}-\tfrac{\hbar}{2}+\cO(q)$. We emphasize that the Bethe wavefunctions $S_I(\sigma^J)$ can be found directly from the gauge theory computation without solving any Bethe equations.


\section{Quantum Mechanical Description}
\label{sec:qm}

\begin{figure}[htp]
\centering
\includegraphics[height=3.5cm]{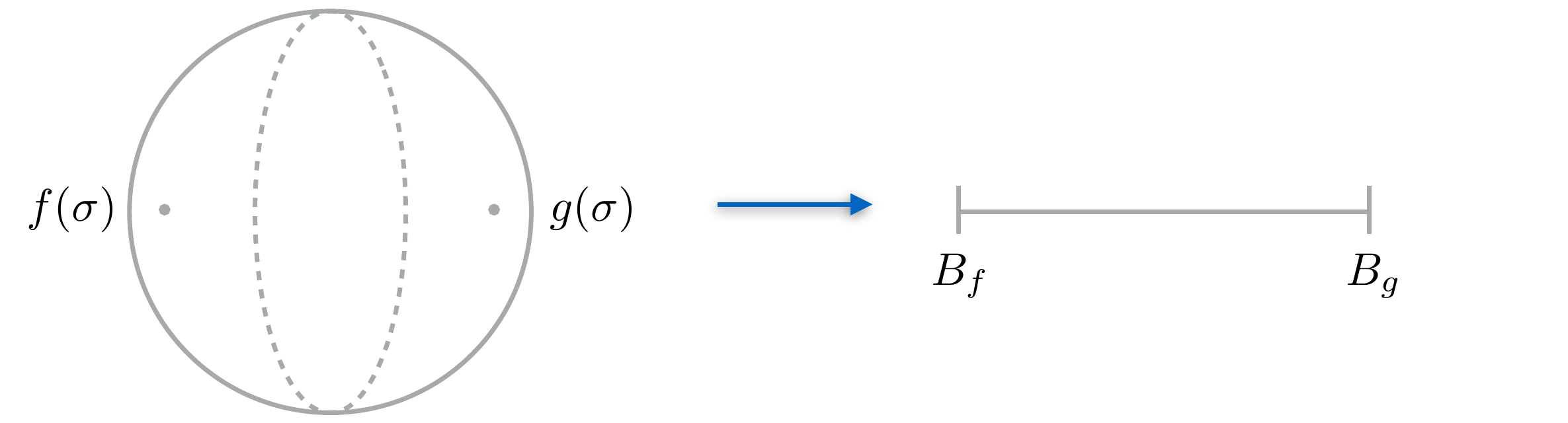}
\caption{Quantum Mechanics Limit}
\label{fig:qmlimit}
\end{figure}

In this section, we provide an alternative viewpoint on $A$-model computations on $\mathbb{CP}^1$ in terms of supersymmetric quantum mechanics. We replace $\mathbb{CP}^1$ by a long cylinder capped off by $A$-twisted cigars. In the cylindrical region, the theory preserves 2d $\cN=(2,2)$ supersymmetry and we can reduce on $S^1$ to obtain an effective $\cN=4$ supersymmetric quantum mechanics. The capped regions  become boundary conditions in the supersymmetric quantum mechanics preserving the supercharges $\Q_-,\overline \Q_+$. This setup is summarized in figure~\ref{fig:qmlimit}.

In principle, we can find a finite dimensional $\cN=4$ supersymmetric quantum mechanics for each flux sector $n \in \mathbb{Z}$ individually. Here we restrict ourselves to a description to the zero flux sector $n=0$, or equivalently the limit $q \to 0$. As we have emphasized in the introduction, much of the representation theoretic apparatus of the algebraic Bethe ansatz does not depend on the choice of quasi-periodic boundary condition specified by $q$, and should therefore have a description in this supersymmetric quantum mechanics.

After a description of the $\cN=4$ supersymmetric quantum mechanics and a general description of how to translate insertions of twisted chiral operators to boundary conditions, we will provide two explicit constructions of the boundary conditions that arise from insertions $S^{(\pi)}_I(\vec\sigma)$. The first involves a combination of Neumann or Dirichlet boundary conditions coupled to boundary degrees of freedom. The second involves the notion of a thimble boundary condition.

\subsection{\texorpdfstring{$\cN=4$}{} Quantum Mechanics}

Let us first consider the $\cN=4$ supersymmetric quantum mechanics obtained from the zero flux sector $n=0$ of our 2d $\cN=(2,2)$ theory, which is obtained by plain dimensional reduction of section~\ref{sec:model} on a circle. In the absence of mass deformations, the supersymmetric quantum mechanics has $U(1)_V \times SU(2)_A$ R-symmetry with $U(1)_A \subset SU(2)_A$ as the Cartan subalgebra. We denote the euclidean coordinate of the supersymmetric quantum mechanics by $\tau$. We refer the reader to appendix \ref{app:SQM} for further details on our conventions for supersymmetric quantum mechanics.

The $U(k)$ vectormultiplet now contains scalar fields $\sigma^A$ transforming in a triplet of the $SU(2)_A$ R-symmetry. The complex combination $\sigma = \sigma^1 + i \sigma^2$ is the two-dimensional scalar and $\sigma^3$ arises from the integral of the two-dimensional gauge field around $S^1$. The fermion gauginos $\lambda$ now transform as a spinor of $SU(2)_A$. The lagrangian for the vectormultiplet $(A_\tau,\sigma^A,\lambda,D)$ is
\be
\mathcal{L}_{\rm vec} = \frac{1}{2e^2}{\rm Tr}\left[- D_\tau \sigma^A D_\tau \sigma^A -\bar\lambda D_\tau\lambda + D^2 + \frac{1}{2}[\sigma^A,\sigma^B]^2 + \bar\lambda \tau^A [\sigma^A,\lambda] \right] \ ,
\ee
where $\tau^A$ denote the $SU(2)_A$ gamma matrices and we have suppressed contractions over $SU(2)_A$ spinor indices. A chiral multiplet consists of a complex scalar $\phi$ and fermions $\psi$ transforming as a spinor of $SU(2)_A$ with lagrangian
\be
\mathcal{L}_{\rm chiral} = -| D_\tau\phi|^2 - \bar\psi D_\tau\psi - |\sigma^A\phi|^2 + \bar\phi D \phi + \bar\psi \tau^A \sigma^A  \psi  + i\bar\psi \bar\lambda\phi + i\bar\phi\lambda\psi +|F|^2 \ .  
\ee
We have $N$ fundamental chiral multiplets $X$, $N$ anti-fundamental chiral multiplets $Y$, and an adjoint chiral multiplet $\Phi$. There is also a cubic superpotential $W={\rm Tr}( \Phi XY)$. 

The supersymmetric quantum mechanics has a vacuum manifold determined by solutions to the equations
\begin{gather}
\label{eq:D-term-qm}
 \mu_\R - r \, \mathbb{1}  = 0 \,,\qquad [\sigma^A , \sigma^B] = 0 \,,\\
\label{eq:F-term-qm}
\Phi \cdot X = 0 \,,\qquad Y \cdot \Phi = 0 \,,\qquad X\cdot Y = 0 \,, \\
\sigma^A \cdot X = 0\,, \qquad - Y \cdot \sigma^A = 0\,, \qquad [\sigma^A,\Phi] = 0 \,,
\label{eq:vector=qm}
\end{gather}
which coincide with configurations annihilated by all four supercharges. As in the two-dimensional theory, with $r>0$ solutions are forced to have $\sigma^A=0$ and the vacuum manifold is $\vacman = T^*G(k,N)$.

The supersymmetric quantum mechanics has flavour symmetry $G_f = PSU(N) \times U(1)_\hbar$. We can turn on $SU(2)_A$ triplets of mass parameters $(m^A_1,\ldots,m^A_N,\hbar^A)$ of mass parameters by coupling to a background vectormultiplet with a vacuum expectation value for $\sigma^A$ in $T_f$. The $\sigma$ components of these mass parameters are the complex masses $(m_1,\ldots,m_N,\hbar)$ introduced in two dimensions. In section~\ref{sec:thimbles}, we will also want to turn on real mass parameters $(m_1^3,\ldots,m_N^3)$ which arise from turning on background holonomy for the $PSU(N)$ flavour symmetry around the $S^1$. 

In the presence of such mass parameters, equations~\eqref{eq:vector=qm} are deformed to
\begin{gather}
(\sigma_a -m_j+\tfrac{\hbar}{2} ) X^a{}_j = 0\,, \qquad (- \sigma_a -m_j+\tfrac{\hbar}{2} ) Y^j{}_a = 0\,, \qquad
(\sigma_a - \sigma_b +\hbar)\Phi^a{}_b = 0\,, \\
(\sigma^3_a -m^3_j) X^a{}_j = 0\,, \qquad (- \sigma^3_a -m^3_j ) Y^j{}_a = 0\,, \qquad
(\sigma^3_a - \sigma^3_b)\Phi^a{}_b = 0\,.
\end{gather}
For generic values of the mass parameters, this lifts the vacuum manifold to the isolated fixed points of the $T_f$ action on $\vacman$.

\subsection{Empty Boundary Condition}

Before constructing boundary conditions in the supersymmetric quantum mechanics that correspond to the inserting $S_I^{(\pi)}(\sigma)$ at the tip of the cigar, we first consider the empty boundary condition associated to a cigar without any insertion. Our discussion of such boundary conditions has much in common with the description of $B$-type boundary conditions in 2d $\cN=(2,2)$ gauge theories~\cite{Herbst:2008jq,Hori:2013ika,Honda:2013uca}.

The appropriate boundary condition for the vectormultiplet can be determined as follows. First, as we are considering the sector with vanishing flux and $\sigma^3 \propto \int_{S^1} A$, we should impose the Dirichlet boundary conditions $\sigma^3|=0$. The remaining boundary conditions,
\be
A_\tau| = 0\,,  \qquad \sigma^3| = 0\,, \qquad D_\tau\sigma| = 0\,,  \qquad \lambda_1|=\bar\lambda_1|=0 \, ,
\ee
are determined by preserving the 0d $\cN=(0,2)$ supersymmetry algebra generated by $\Q_1=\Q_-$ and $\overline{\Q}_1=\overline{\Q}_+$ at the boundary.

Let us now consider the boundary conditions for the chiral multiplets. A 1d $\cN=4$ chiral multiplet can be decomposed into a chiral multiplet $(\phi,\psi_2)$ and a Fermi multiplet $\psi_1$ with superpotential $E_{\psi}(\phi) = D_\tau\phi$ in terms of the boundary $\cN=(0,2)$ supersymmetry algebra generated by $\Q_-$ and $\overline{\Q}_+$. A basic boundary condition therefore involves Neumann for the chiral component and Dirichlet for the Fermi component,
\be
	D_\tau \phi = 0|\,,  \qquad \psi_1| = 0\,, \qquad D_\tau\psi_2 | = 0 \, .
\ee
The components $(\phi,\psi_2)$ transform as a $\mathcal{N}=(0,2)$ chiral multiplet at the boundary.
We call this a `Neumann' boundary condition. In the presence of a bulk superpotential, this boundary condition must be supplemented by a choice of matrix factorization.

The boundary condition corresponding to the empty cigar can now be described as follows. We first impose Neumann boundary conditions for $X$, $Y$ and $\Phi$. We then couple to a 0d Fermi multiplet $\eta$ with the same charges as $\Phi$ and boundary superpotentials 
\be
E_\eta = \Phi| \, , \quad J_\eta=X\cdot Y| \, .
\ee
This provides a matrix factorization of the bulk superpotential, $W| = E_\eta \cdot J_\eta$. Note that the boundary superpotential $E_\eta=\Phi|$ effectively modifies the boundary condition for $\Phi$ from Neumann to Dirichlet. This is compatible with the zero mode analysis in section~\ref{sec:orbifold}. In particular, the chiral fields $X(z)$ and $Y(z)$ were non-zero at $z=0$ corresponding to a Neumann boundary condition, whereas $\Phi(0) = 0$ reproducing a Dirichlet boundary condition.

The `empty' boundary condition is compatible with any vacuum configuration in $\vacman$.
It will therefore flow to Neumann boundary conditions in the supersymmetric sigma model supported on the whole of $\vacman$. In particular, the partition function on an interval with the empty boundary condition at each end reproduces the $n=0$ contribution to the $\mathbb{CP}^1$ partition function,
\be
\la 1 \ra_{1d}=(-1)^{k^2}\int_\gamma \frac{d^k\sigma}{k!} \frac{\prod\limits_{a \neq b}(\sigma_a-\sigma_b)\prod\limits_{a,b=1}^k(\sigma_a-\sigma_b-\hbar)}{\prod\limits_{a=1}^k \prod\limits_{i=1}^N(\sigma_a-m_i+\tfrac{\hbar}{2} )(-\sigma_a+m_i+\tfrac{\hbar}{2} )}  \, ,
\ee
which is the equivariant integral of `$1$' over $\vacman$. From this perspective, the denominators arise from the $\cN=(0,2)$ chiral multiplets $X,Y$. The contributions coming from two Fermi multiplets $\eta$ at two boundaries cancel the contribution from the chiral multiplet $\Phi$ and provide the factor $(-1)^{k^2}\prod_{a,b}(\sigma_a-\sigma_b-\hbar)$ in the numerator. The remaining numerator factor $\prod_{a \neq b}(\sigma_a-\sigma_b)$ is the vector multiplet contribution. The contour $\gamma$ is chosen by following the Jeffrey-Kirwan prescription.

\subsection{Stable Boundary Conditions}
\label{sec:stable-bc}

We now consider the class of boundary conditions that arise from inserting $S_I(\sigma)$ at the tip of the cigar. Flowing to a supersymmetric sigma model to the vacuum manifold $\vacman$, such boundary conditions are supported on holomorphic lagrangian submanifolds in $\vacman$ that are fixed by $T_f$. In this section, we provide an elementary description of these boundary conditions in the supersymmetric gauge theory by coupling the empty boundary condition to additional boundary degrees of freedom. The construction is similar in spirit to the exceptional Dirichlet boundary conditions introduced in~\cite{Bullimore:2016nji}.

Let us first consider the abelian case. Following the arguments of section~\ref{sec:twisted-chiral}, let us first consider the boundary condition obtained by inserting the homogeneous polynomial $\sigma - m_j +\frac{\hbar}{2}$ at the tip of the cigar,
corresponding to the equivariant weight of the coordinate $X_j$. In the vanishing flux sector, this is a polynomial representative of the equivariant cohomology class Poincar\'e dual to $\{X_j = 0\} \subset T^*\mathbb{CP}^{N-1}$.

This boundary condition is described in the supersymmetric quantum mechanics by coupling the empty boundary condition to a 0d $\mathcal{N}=(0,2)$ Fermi multiplet $\chi_j$ with the same charges as $X_j$ with boundary superpotentials
\be
	E_{\chi_j} = X_j \ , \quad J_{\chi_j} = 0 \ .
\ee
The 0d $\cN=(0,2)$ chiral multiplet part of $X_j$ will receive a mass from the superpotential at the boundary modifying the boundary condition for $X_j$ from Neumann to Dirichlet. This boundary condition is therefore supported on the subspace $\{X_j=0\} \subset T^* \mathbb{CP}^{N-1}$ of the vacuum manifold. In the computation of partition functions, the boundary Fermi multiplet $\chi_j$ provides an additional contribution
\be
	\sigma -m_j + \frac{\hbar}{2} \,,
\ee
in the numerator of the integrand. This reproduces the effect of inserting the twisted chiral operator $f(\sigma)=\sigma -m_j + \frac{\hbar}{2}$ at the tip of the cigar. 

It is now straightforward to write down the boundary condition corresponding to inserting the twisted chiral operator $S_I(\sigma)$ with $I=\{i\}$ in abelian theories. We deform the empty boundary condition by introducing $(N-1)$ boundary Fermi multiplets $\chi_j$ for $j  \neq i$ with the boundary superpotentials
\be
	E_{\chi_j} = \begin{cases} X_j & \mathrm{if} \quad j<i \\ Y_j & \mathrm{if} \quad j>i \end{cases}\, , \qquad 
	J_{\chi_i} = 0 \ .
\ee
The charges of the Fermi multiplets are uniquely determined by their superpotentials. The boundary condition is now supported on the subspace
\be
	X_j = 0 \quad {\rm for} \ \ j < I \ , \qquad Y_j = 0 \quad {\rm for} \ \ j > I\,, 
\ee
of the vacuum manifold $\vacman$. A partition function with this boundary condition will include an additional contribution compared to the empty boundary condition,
\be
	\prod_{i=1}^{I-1}\left(\sigma-m_i+\tfrac{\hbar}{2}\right) \prod_{i=I+1}^N\left(-\sigma+m_i+\tfrac{\hbar}{2}\right) \ ,
\ee
from the additional contributions of the Fermi multiplets $\chi_j$. Therefore, this boundary condition reproduced the computation with $S_I(\sigma)$ inserted. Note that due to the symmetry $(\chi_j,E_j,J_j) \leftrightarrow (\bar\chi_j, J_j,E_j)$, we could equivalently have coupled to boundary fermi multiplets with $J$-type superpotentials.

This boundary condition can be easily generalized to non-abelian theories in a manner consistent with the zero mode analysis of section~\ref{sec:orbifold}. The boundary condition for $S_I(\vec\sigma)$ with $I = \{I_1,\ldots,I_k\}$ should be supported on the subspace of $\vacman$ defined by 
\be\label{eq:stable-basis-SQCD}
	X^a_{\ j} = 0 \quad {\rm for} \ \ j < I_a \ , \qquad Y_{\ \ a}^{j} = 0 \quad {\rm for} \ \ j > I_a \ .
\ee
These constraints can be implemented by introducing $k(N-1)$ Fermi multiplets denoted by $\chi^a{}_j$ for $j<I_a$ and $\chi^j{}_a$ for $j>I_a$ with superpotentials
\be
	(E_{\chi})^a_{\ j} = X^a_{\ j} \quad {\rm for} \ \ j < I_a \ , \qquad (E_{\chi})_{\  a}^j = Y_{\ \ a}^j \quad {\rm for} \ \ j > I_a \ ,
\ee
together with $J_\chi =0$. 

Recalling that $I_a \leq I_b$ if and only if $a \leq b$, these constraints imply that the complex moment map for the gauge symmetry is upper triangular at the boundary,
\be
\left. (X \cdot Y )^a{}_b \right|  =  \begin{cases}
\displaystyle\sum\limits_{I_a < j < I_b} X^a{}_j Y^j{}_b |\,, & \mathrm{if} \quad a \leq b\,, \\
0\,, & \mathrm{if} \quad a > b\,.
\end{cases}
\ee
Therefore the matrix factorization requires that we introduce only $\frac{k(k+1)}{2}$ Fermi multiplets at the boundary, with components $\eta^a{}_b$ for $a \geq b$, with boundary superpotentials $(J_\eta)^a{}_b = X\cdot Y|$ and $(E_\eta)^a{}_b=\Phi^a{}_b|$. This leads to additional bosonic fluctuations at the boundary from the components $\Phi^a{}_b$ with $a<b$.

In addition, the boundary condition breaks complex gauge transformations $G_\C = GL(k)$ to the Borel subgroup $B$ of upper triangular transformations. There are therefore additional bosonic fluctuations at the boundary parametrizing $\mathbb{F}_k = G_\C / B$, which is complex version of the breaking of $U(k)$ to its maximal torus described in section~\ref{sec:orbifold}. As mentioned there, these fluctuations combine with those of $\Phi^a{}_b|$ for $a<b$ to form the cotangent bundle $T^*\mathbb{F}_k$, whose hyper-K\"ahler structure is a reflection of the fact that the boundary condition preserves a $\cN=(0,4)$ supersymmetry in the limit $\hbar \to 0$.

In the computation of interval partition functions, this boundary condition leads to an additional contribution compared to the empty boundary condition,
\be
S_I(\sigma) = 	\frac{\prod\limits_{a=1}^k\left(\prod\limits_{i=1}^{I_a-1}(\sigma_a-m_i+\frac{\hbar}{2})\prod\limits_{I_a+1}^N(-\sigma_a+m_i+\frac{\hbar}{2})\right)}{\prod\limits_{a<b}(\sigma_a - \sigma_b)(\sigma_a-\sigma_b-\hbar)} \ .
\ee
Here, the numerator factors are the contributions from the boundary Fermi multiplets $\chi^a{}_j$ and $\chi^j{}_a$. The denominator factors arise from additional bosonic zero modes of $\Phi^a{}_b|$ for $a<b$ and from the breaking of the gauge symmetry. As mentioned in section~\ref{sec:orbifold}, the denominator factors combine to form an equivariant weight of the tangent bundle at a fixed point of $T^*\mathbb{F}_k$, and symmetrizing over $\sigma_1,\ldots,\sigma_k$, the result can be interpreted as an equivariant integral over this boundary moduli space.

\subsection{Thimble Boundary Conditions}
\label{sec:thimbles}

In this section, we turn on real mass parameters $(m^3_1,\dots,m^3_N)$ valued in the Cartan subalgebra of the $PSU(N)$ flavour symmetry. These parameters arise from a background flavour holonomy around $S^1$ in the two-dimensional setup. In the presence of real mass parameters, there is a natural class of boundary conditions for the supersymmetric quantum mechanics preserving $\Q_-$, $\overline \Q_+$, which are analogous to thimble boundary conditions in 2d $\cN=(2,2)$ theories~\cite{Hori:2000ck,Gaiotto:2015aoa,Gaiotto:2015zna}. We will need a slight generalization of the standard notion appropriate for theories with multiple isolated vacua connected by gradient flows~\cite{Bullimore:2016nji}. We will adapt this construction here to the context of 1d $\cN=4$ supersymmetric quantum mechanics. We expect this to provide a physical counterpart to the construction of stable envelopes~\cite{Maulik:2012wi}.

In the presence of real mass parameters $\{m^3_1,\dots,m^3_N\}$, the configurations of the supersymmetric quantum mechanics preserving $\Q_-$, $\overline \Q_+$ are
\bea
\mu_\R - r \mathbb{1}&=- \frac{1}{g^2} D_{\tau} \sigma^3  \,, \\
D_\tau X   =   - \sigma^3 \cdot X  +  X \cdot m^3 \,,\qquad
D_\tau Y  &=  Y \cdot \sigma^3  - m^3 \cdot Y\,,\qquad
D_{\tau} \Phi   =  -  [\sigma^3,\Phi]\,.
\eea
These equations can be reformulated as the gradient flow equations
\be
D_\tau \mathbb{X} =  - g^{\mathbb{X} \mathbb{X}^\dagger } \frac{\delta h}{ \delta \mathbb{X}^\dagger} \,,
\ee
where
\be
h =  \int_{\gamma} \sigma^3 \cdot (\mu_\R - r \mathbb{1} )  - m^3 \cdot \mu_{\R,m}\,,
\ee
and $\mu_\R$ and $\mu_{\R,m}$ denote the moment maps for the gauge and flavour symmetry respectively. The notation $\mathbb{X}$ refers to one of the fields $\{ X,Y,\Phi,\sigma^3\}$ and $g^{\mathbb{X} \mathbb{X}^\dagger }$ is the inverse metric on the space of fields inherited from the Lagrangian. 

Note that inside of gauge invariant combinations of the fields $X$ and $Y$, we have
\bea
\partial_\tau X  & =   X \cdot m^3\,, \qquad\qquad
\partial_\tau Y  & = - m^3 \cdot Y \, ,
\label{eq:vacman-gradient-flow}
\eea
and therefore gauge invariant combinations of $X$ and $Y$ will grow or decay along the $\tau$ direction according to their charge under the flavour transformation generated by $m^3$. This is gradient flow on the  vacuum manifold $\vacman$ for the Morse function $h_m =  m^3 \cdot \mu_{\R,m}$.

A preliminary definition of a left thimble boundary condition $B_I$ on $\tau \geq 0$ can now be given as follows: it is a boundary that is equivalent for the computation of correlation functions preserving $\Q_-$, $\overline \Q_+$ to the placing theory on $\tau \geq - \infty$ with a fixed isolated vacuum $v_I$ at $\tau \to - \infty$. The support of such a boundary condition is in the first instance the submanifold of points in $\vacman$ that can be reached by an infinite gradient flow from the vacuum $v_I$ at $\tau \to - \infty$. This submanifold clearly depend on the ordering of the real mass parameters. Suppose that the real masses are ordered as $m^3_{\pi(1)}< \cdots< m^3_{\pi(N)}$ for some permutation $\pi$, then we denote this submanifold by $\vacman^{(\pi)}_I$.

However, in passing from $\tau = - \infty$ to $\tau = 0$, we will need to allow for a sequence of domain walls preserving $\Q_-$, $\overline \Q_+$ that interpolate between different isolated vacua connected by gradient flows. In order to formalize this notion, we can introduce a partial ordering on the set of isolated vacua $\{v_I\}$ depending on the permutation $\pi$, by the requirement
\be
\text{there exists a $\pi$-gradient flow $v_I \to v_J$} \quad \Rightarrow \quad v_I <_\pi v_J \, , 
\ee
and extending transitively, namely if $v_I<_\pi v_J$ and $v_J<_\pi v_K$ then also $v_I<_\pi v_K$. 
Allowing for sequences of domain walls interpolating between vacua, the support of a left thimble boundary condition $B_I$ is
\be
\vacman_{B_I}^{(\pi)} = \bigcup _{v_I \leq_\pi v_J} \vacman_{J}^{(\pi)} \, .
\ee
We claim that the thimble boundary condition $B_I$ in the presence of mass parameters ordered by the permutation $\pi$ corresponds to boundary condition constructed in section \ref{sec:stable-bc} associated to inserting $S^{(\pi)}_I(\sigma)$.

Let us illustrate this construction with an abelian example: $k=1$ and $N=2$ with $\vacman = T^*\mathbb{CP}^1$ - see table~\ref{tab:charges}. Introducing a real mass parameter $m^3$ for the $U(1)_m$ flavour symmetry, we have
\be
h_m = m^3(-|X_1|^2+|X_2|^2+|Y_1|^2-|Y_2|^2) \, .
\ee
Note that in the graphical representation of $\vacman = T^*\mathbb{CP}^1$ in figure~\ref{fig:vac-sqed}, the Morse function is proportional to the coordinate along the horizontal axis. The gradient flows are therefore straightforward to understand in this graphical representation.

\begin{figure}[htp]
\centering
\includegraphics[height=5.5cm]{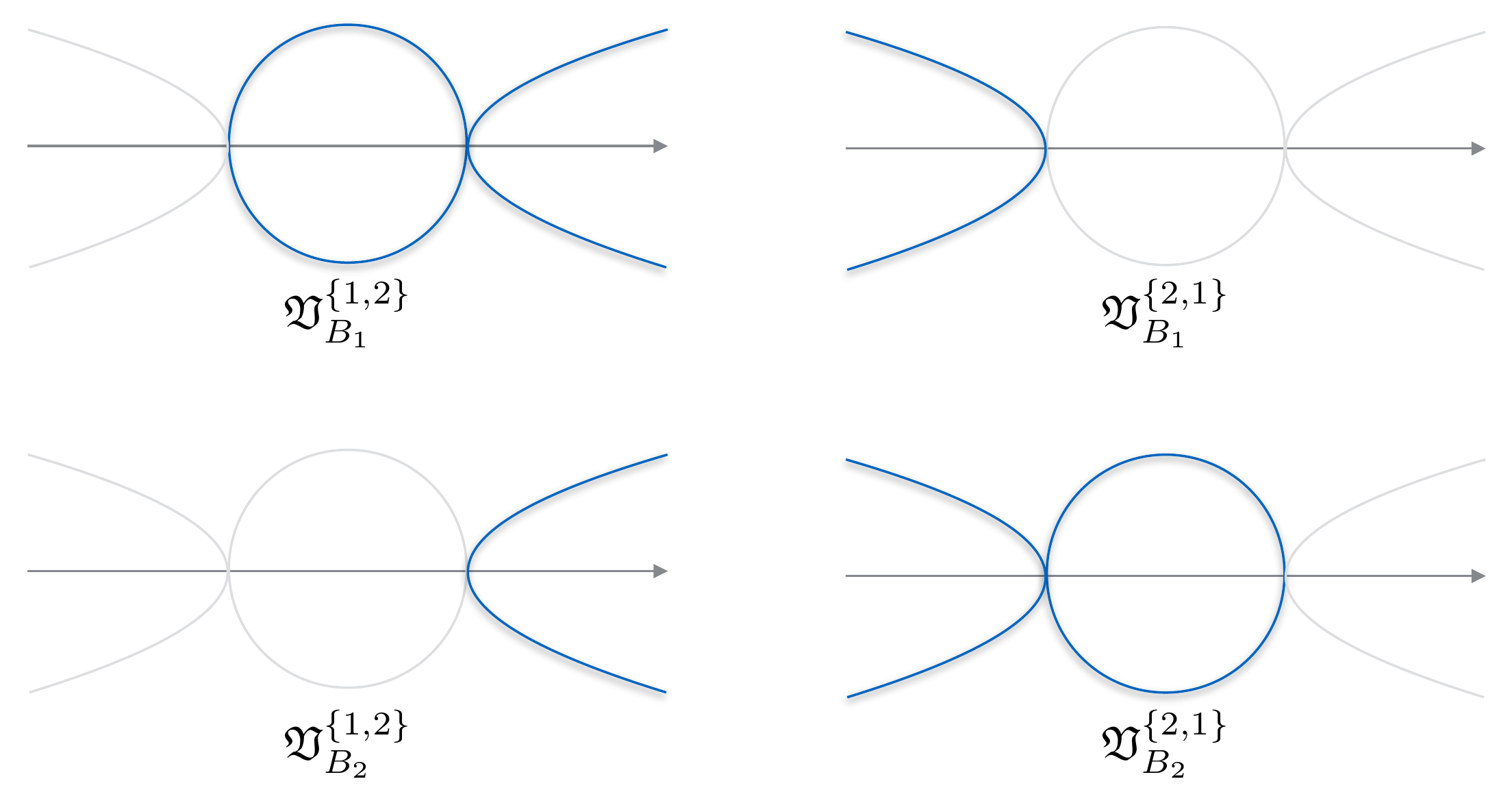}
\caption{The support of thimble boundary conditions generated by vacua $v_1$ and $v_2$ for permutations $\pi=\{1,2\}$ ($m^3<0$) and $\pi = \{2,1\}$ ($m^3>0$).}
\label{fig:grad2}
\end{figure}

The permutation $\pi=\{1,2\}$ corresponds to $m^3<0$ and the permutation $\pi = \{2,1\}$ corresponds to $m^3>0$. In the graphical representation, the direction of flow for increasing $\tau$ is from left to right for $\pi=\{1,2\}$ and right to left for $\pi = \{2,1\}$. The vacua are therefore ordered such that $v_1 <_\pi v_2$ for $\pi = \{1,2\}$ and $v_2 <_\pi v_1$ for $\pi = \{2,1\}$. First, it is straightforward to see that 
\bea
\vacman_1^{\{1,2\}} & = \mathbb{CP}^1 - \{v_2\}\,, \qquad && \vacman_1^{\{2,1\}} = F_1 \,, \\
\vacman_2^{\{1,2\}} &= F_2 \,,\qquad &&  \vacman_2^{\{2,1\}} = \mathbb{CP}^1 - \{v_1\}\,,
\eea
where $F_i$ denotes the fiber of $\vacman = T^*\mathbb{CP}^1$ at the fixed point $v_i$. We therefore generate thimble boundary conditions with support
\bea
\vacman_{B_1}^{\{1,2\}} & = \vacman_{1}^{\{1,2\}} \cup \vacman_{2}^{\{1,2\}} = \mathbb{CP}^1 \cup F_2 \,,\qquad && \vacman_{B_1}^{\{2,1\}} = \vacman_{2}^{\{2,1\}} = F_1\,, \\
\vacman_{B_2}^{\{1,2\}} & = \vacman_{2}^{\{1,2\}} = F_2 \,, \qquad && \vacman_{B_2}^{\{2,1\}} = \vacman_{2}^{\{2,1\}} \cup \vacman_{1}^{\{2,1\}} = \mathbb{CP}^1 \cup F_1 \, .
\eea
which are illustrated in figure~\ref{fig:grad2}. The supports of these boundary conditions clearly coincide with those obtained from the cigar with insertions of $S^{\{1,2\}}_{1}(\sigma)$ and $S^{\{1,2\}}_{2}(\sigma)$ for $\pi=\{1,2\}$, and $S^{\{2,1\}}_{1}(\sigma)$ and $S^{\{2,1\}}_{2}(\sigma)$ for $\pi=\{2,1\}$.

Finally, let us attempt to make a general statement. We expect that the left thimble boundary condition $B_I$ generated by real mass parameters $m^3_{\pi(1)} < \cdots m^3_{\pi(N)}$ is equivalent for the purpose of computing correlation functions preserving $\Q_-$, $\overline \Q_+$ to the boundary condition in section~\ref{sec:stable-bc} corresponding to $S^{(\pi)}_I(\sigma)$. Since $\tau \to -\tau$ transforms the gradient flow equations in the same way as $m^3_j \to - m^3_j$, the right thimble boundary condition for the same mass parameters reproduces the function $S^{(\iota \cdot \pi)}_I(\sigma)$, where $\iota : \{1,\ldots,N\} \to \{N,\ldots,1\}$ is the longest permutation.

\section{The R-matrix}
\label{sec:r-matrix}

In this section we return to studying the Heisenberg spin chain and the question of how the $R$-matrix arises in the study of supersymmetric gauge theory. In particular, we will examine the two-point functions of stable basis elements $S_{I}^{(\pi)}(\vec\sigma)$ in the $A$-twisted supersymmetric gauge theory on the sphere. These correlation functions are in fact independent of $q$ and can therefore be interpreted as the partition function of a supersymmetric quantum mechanics of section~\ref{sec:qm} on a interval with thimble boundary conditions at either end.

\subsection{Orthonormality of Stable Basis}

Let us first consider the two-point correlation functions of basis elements $S_I^{(\pi)}(\vec\sigma)$ at $\{+\}$ and $S_J^{(\iota \cdot \pi)}(\vec\sigma)$ at $\{-\}$ where $\iota : \{1,\ldots,N\} \to \{N,\ldots,1\}$ is the longest permutation. This correlation functions are independent of $q$ and evaluate to
\be
\la S_I^{(\pi)}(\vec\sigma) S_J^{(\iota \cdot \pi)}(\vec\sigma) \ra_{S^2} = \delta_{I,J} \, .
\ee
The appearance of the reflection $\iota$ here is natural from the orbifold construction. The orbifold construction at $\{+\}$ producing $S_I^{(\pi)}(\vec\sigma)$ has flavour holonomy $(g_F)^i{}_j = \omega^{\pi(i)-1}\delta^i{}_j$. This is compatible with turning on flavour holonomy in a neighbourhood of $\{+\}$ of the form $e^{i\theta_i}\delta^i{}_j$ such that $\theta_{\pi(1)}<\cdots \theta_{\pi(N)}$. Translating to $\{-\}$ without allowing these holonomy eigenvalues to cross, the compatible orbifold construction leads to functions $S^{(\iota \cdot \pi)}_I(\vec\sigma)$. Similarly in the supersymmetric quantum mechanics, introducing constant real masses ordered such that $m^3_{\pi(1)} < \cdots <m^3_{\pi(N)}$ leads to left thimble boundary conditions generating $S^{(\pi)}_I(\vec\sigma)$ and right thimble boundary conditions generating $S^{(\iota \cdot \pi)}_I(\vec\sigma)$.

This observation motivates to define an inner product,
\be
\label{inner.SC}
\langle f(\vec\sigma) , g(\vec\sigma)\rangle= \langle f(\vec\sigma),\overline{g(\vec\sigma)}\rangle_{S^2}\,,
\ee
where the conjugation sends $\sigma_a\to-\sigma_a,m_i\to-m_i$. In particular, $\overline{S^{(\pi)}_I(\vec\sigma)} = S^{(\iota \cdot \pi)}_I(\vec\sigma)$ and therefore the stable basis elements for a given permutation $\pi$ are orthonormal,
\be
\la S_I^{(\pi)}(\vec\sigma)  , S_J^{(\pi)}(\vec\sigma) \ra = \delta_{I,J} \, .
\label{eq:orth}
\ee
By construction, this inner product depends only on the functions $f(\vec\sigma)$ and $g(\vec\sigma)$ modulo the twisted chiral ring relations. Under the correspondence outlined in the introduction~\ref{sec:intro}, this corresponds to the inner product on the spin chain Hilbert space\footnote{In the supersymmetric gauge theory, we have allowed the parameters $(\sigma_1,\ldots,\sigma_k)$, and $(m_1,\ldots,m_N,\hbar)$ to be complex. In order to recover the honest inner product on the spin chain Hilbert space, we would need to specify certain reality conditions on these parameters such that they map exactly onto the corresponding spin chain parameters.},
\begin{equation}\label{Hilbert.space.perm}
\mathcal{V}^{(\pi)}=\mathbb{C}^2_{m_{\pi(1)}} \otimes \mathbb{C}^2_{m_{\pi(2)}} \otimes  \ldots \otimes \mathbb{C}^2_{m_{\pi(N)}}\,,
\end{equation} 
with sites ordered according to the permutation $\pi$.
In particular, equation~\eqref{eq:orth} corresponds to the orthonormality of the up-down basis of the spin chain, $\la I|J\ra = \delta_{I,J}$.

\subsection{R-matrix from Janus Interface}

The natural next step is to consider the inner product of stable basis elements for different permutations $\pi$ and $\pi'$. These correlation functions are again independent of $q$. Let us therefore consider this problem from the perspective of the supersymmetric quantum mechanics of section~\ref{sec:qm}. 

Consistency of such a correlation function requires that the mass parameters $(m_1^3,\ldots,m_N^3)$ vary as a function of $\tau$ across an interval, such that they are ordered by the permutation $\pi$ at the left boundary and by $\pi'$ at the right boundary. This is an exact deformation and therefore correlation functions do not depend on the particular profile of this variation. In particular, we can say that such a correlation function requires the presence of a `Janus interface' $\cJ_{\pi,\pi'}$ for the real mass parameters. The inner product
\be
\la S_I^{(\pi)}(\vec\sigma)  , S_J^{(\pi')}(\vec\sigma) \ra \,,
\ee
is computing the partition function of the supersymmetric quantum mechanics on a interval with the interface $\cJ_{\pi,\pi'}$ between thimble boundary conditions generated by the vacua $v_I$ on the left and $v_J$ on the right.

\begin{figure}[ht!]
\centering
\includegraphics[height=3cm]{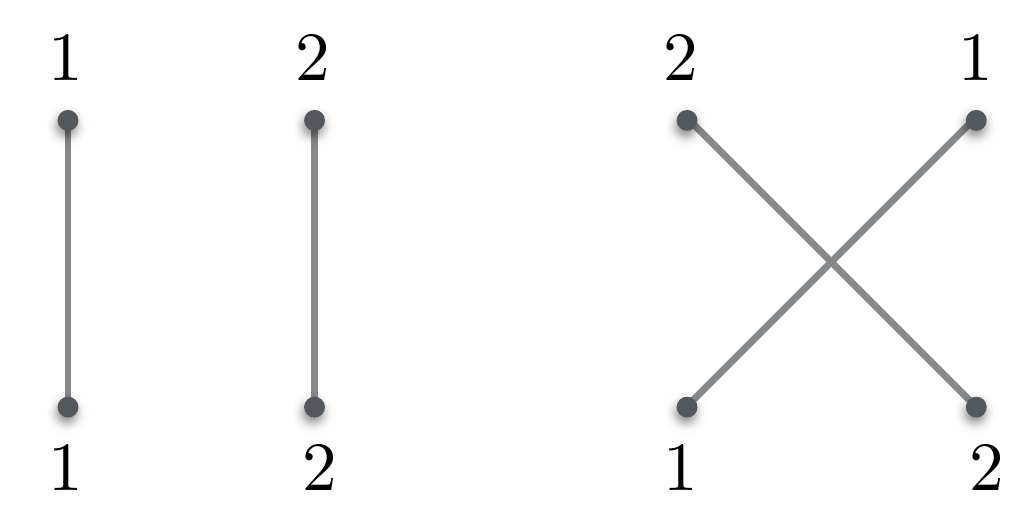}
\caption{Graphical representation of correlation functions producing identity and R-matrix for a spin chain of length $N=2$.}
\label{fig:n2}
\end{figure}

In order to develop the connection between such correlation functions and spin chain quantities, let us consider the simplest example corresponding to a spin chain of length $N = 2$. In that case we have two distinct permutations, $\{1,2\}$ and $\{2,1\}$, and three supersymmetric theories with $k = 0$, $1$ and $2$. The correlation functions 
\be
\langle S_I^{\{2,1\}}(\vec\sigma) , S^{\{1,2\}}_J(\vec\sigma)\rangle\,,
\ee
are straightforward to evaluate explicitly. The result is summarized in the following table
\begin{equation}
\begin{tabular}{c|c|cc|c}
&$S_{\{ \}}^{\{1,2\}}$&$S_{\{ 1 \}}^{\{1,2\}}$&$S_{\{2\}}^{\{1,2\}}$&$S_{\{ 1,2 \}}^{\{1,2\}}$\\\hline
$S_{\{\}}^{\{2,1\}}$&1&0&0&0\\\hline
$S_{\{1\}}^{\{2,1\}}$&0&$\displaystyle\frac{m_1-m_2}{m_2-m_1+\hbar}$&$\displaystyle\frac{\hbar}{m_2-m_1+\hbar}$&0\\
$S_{\{2\}}^{\{2,1\}}$&0&$\displaystyle\frac{\hbar}{m_2-m_1+\hbar}$&$\displaystyle\frac{m_1-m_2}{m_2-m_1+\hbar}$&0\\\hline
$S_{\{1,2\}}^{\{2,1\}}$&0&0&0&1 
\end{tabular} \, .
\end{equation}
One can immediately recognize this table as the matrix elements of the spin chain R-matrix \eqref{Rmatrix.explicit} acting on $\C^2_{m_1} \otimes \C^2_{m_2}$, up to a sign. Summarizing this example, there are two sets of correlation functions
\bea
\langle S_I^{\{1,2\}}(\vec\sigma) , S^{\{1,2\}}_J(\vec\sigma)\rangle & =\delta_{IJ}\,, \\
 \langle S_I^{\{2,1\}}(\vec\sigma) , S^{\{1,2\}}_J(\vec\sigma)\rangle & \sim R_{IJ}(m_1-m_2)\,,
\eea
which can be represented graphically as in figure~\ref{fig:n2}. The first line consists of correlation functions consistent with constant real masses ordered by the permutation $\{1,2\}$, whereas the second line contains correlation functions consistent with the presence of a Janus interface $J_{\{2,1\},\{1,2\}}$.

In order to extend this correspondence to supersymmetric gauge theories with $N>2$, we will need to introduce the notion of a Weyl R-matrix depending on a pair of permutations $\pi_+$, $\pi_-$, 
\be
R^{(\pi_+,\pi_-)}(m_1,\ldots,m_N) : \cV^{(\pi_-)} \longrightarrow \cV^{(\pi_+)} \, .
\ee
In order to define it, let decompose the $\pi_+ \cdot \pi_-^{-1}$ as a convolution of elementary transpositions of two elements $(i\,j)$,
\begin{equation}\label{permutation.decomposition}
\pi_+ \cdot \pi_-^{-1}=(i_1 \, j_1)\cdot (i_2 \, j_2)\cdot \ldots \cdot(i_L\, j_L)\,.
\end{equation} 
We now define
\begin{align}\nonumber
R^{(\pi_+,\pi_-)}(m_1,\ldots,m_N)&:=R_{\pi_-(i_1)\,\pi_-(j_1)}\left(m_{\pi_-(j_1)}-m_{\pi_-(i_1)}\right)\cdot\ldots\\\label{Weyl.Rmatrix}
&\hspace{3cm}\ldots \cdot R_{\pi_-(i_L)\,\pi_-(j_L)}\left(m_{\pi_-(j_L)}-m_{\pi_-(i_L)}\right)\,.
\end{align}
As a consequence of the Yang-Baxter equation \eqref{Yang.Baxter} and the unitarity condition \eqref{unitarity} the Weyl R-matrix is independent of the choice of decomposition into elementary transpositions. 

 We have found that the components of the Weyl R-matrix coincide with the correlation functions
\begin{equation}\label{Weyl.matrix}
R^{(\pi_+,\pi_-)}_{IJ}(m_1,\ldots,m_N)=\mathcal{N}_{IJ}^{(\pi_+,\pi_-)} \times \langle S^{(\pi_+)}_I(\vec\sigma) , S^{(\pi_-)}_J(\vec\sigma)\rangle \, ,
\end{equation}
where
\be
\mathcal{N}^{(\pi_+,\pi_-)}_{IJ}=(-1)^{|\pi_+^{-1}(I)|+|\pi_-^{-1}(J)|}\,,
\ee
is a sign. We have performed extensive checks of this relation in numerous examples. In the supersymmetric quantum mechanics setup of section~\ref{sec:qm}, the matrix elements of the Weyl R-matrix are therefore identified with the partition function of a Janus interface $\cJ_{\pi_+,\pi_-}$ between thimble boundary conditions generated by the vacua $v_I$ and $v_J$.

\begin{figure}[ht!]
\centering
\includegraphics[height=3cm]{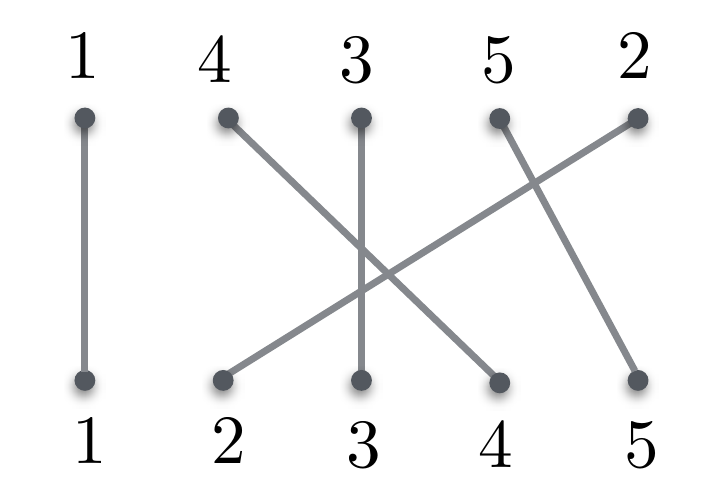}
\caption{Decomposition of the Weyl R-matrix as a product of elementary R-matrices. The Weyl R-matrix is independent of the decomposition as a consequence of the Yang-Baxter equation (figure~\ref{Fig:YangBaxter.equation}) and unitarity (figure~\ref{Fig:unitarity}).} 
\label{fig:factorize}
\end{figure}

We end this section with an example of a Weyl R-matrix. Let us fix the spin chain length $N=5$ and choose permutations $\pi_+=\{ 1,4,3,5,2\}$ and $\pi_-=\{ 1,2,3,4,5\}$. We can decompose the permutation $\pi_+\cdot \pi_-^{-1}=(25)\cdot(34)\cdot(24)\cdot(23)$ and therefore define
\begin{equation}\label{Rmatrix.example}
R^{(\{1,4,3,5,2\},\{ 1,2,3,4,5\})}(m_1,\ldots,m_5)=R_{25}(m_{52})R_{34}(m_{43})R_{24}(m_{42})R_{23}(m_{32})\,,
\end{equation}
which can be straightforwardly computed from the matrix elements of the elementary R-matrices, $R_{ij}(m_{ji})$. This R-matrix can be depicted as in figure~\ref{fig:factorize}. Equivalently, the same matrix elements can be computed from the correlation functions
\begin{equation}
R^{(\{1,4,3,5,2\},\{ 1,2,3,4,5\})}_{IJ}(m_1,\ldots,m_5)=\mathcal{N}^{}_{IJ}\langle S_I^{\{1,4,3,5,2\}}(\vec\sigma),S^{\{1,2,3,4,5\}}_J(\vec\sigma)\rangle\,,
\end{equation}
which can be evaluated by directly performing the contour integral in \eqref{correlation.function}. Both of these computations give the same result.

\subsection{Yang-Baxter Equation}

As explained in the previous section, we have performed extensive checks that the Weyl R-matrix $R^{(\pi_+,\pi_-)}_{IJ}(m_1,\ldots,m_N)$ corresponds to matrix elements of a Janus interface $\cJ_{\pi_+,\pi_-}$ in the supersymmetric quantum mechanics setup of section~\ref{sec:qm}. 

In the same manner that the Weyl R-matrix $R^{(\pi_+,\pi_-)}_{IJ}(m_1,\ldots,m_N)$ is constructed from elementary R-matrices according to a decomposition $\pi_+ \cdot \pi_-^{-1}=(i_1 \, j_1)\cdot \ldots \cdot(i_L\, j_L)$, the Janus interface $\cJ_{\pi_+,\pi_-}$ is a composition of elementary Janus interfaces $\cJ_{ij}$ that interchanges the real masses $m_i^3$ and $m_j^3$,
\be
\label{eq:Janus.Weyl}
\cJ_{\pi_+,\pi_-}= \cJ_{i_1j_1}\cdot \ldots \cdot \cJ_{i_Lj_L}\,.
\ee
This can be understood since deformations of the profile $m_1^3(\tau),\ldots,m_N^3(\tau)$ for the real mass parameters are exact in $\Q_-$, $\overline \Q_+$. We are therefore free to choose a profile consisting of a sequence of `jumps' where pairs of mass parameters $m^3_i$ and $m^3_j$ are interchanged. Each of these jumps can be regarded as an elementary Janus interface $\cJ_{ij}$. Inserting the complete set of states provided by the stable basis $S^{(\pi)}_I(\vec\sigma)$ in between each elementary Janus interface then reproduces the decomposition~\eqref{Weyl.Rmatrix} of the Weyl R-matrix. Equation~\eqref{eq:Janus.Weyl} can therefore be understood as a basis-independent statement of this decomposition.

The fact that the Weyl R-matrix is independent of the choice of decomposition into elementary transpositions followed from the Yang-Baxter equation~\eqref{Yang.Baxter} and the unitarity condition~\eqref{unitarity}. From the perspective of supersymmetric quantum mechanics, this property is guaranteed since different decompositions of a profile $m_1^3(\tau),\ldots,m_N^3(\tau)$ into elementary jumps are related by exact deformations. In particular, we have
\begin{equation}\label{eq.Janus.Yang-Baxter}
\cJ_{ij}\cdot \cJ_{ik}\cdot \cJ_{jk}=\cJ_{jk}\cdot \cJ_{ik}\cdot \cJ_{ij}\,,
\end{equation}
and
\begin{equation}\label{eq.Janus.unitarity}
\cJ_{ij}\cdot \cJ_{ji}= \cI \,,
\end{equation}
where $\cI$ is an identity interface preserving the order of the real mass parameters. This is a basis-independent statement of the Yang-Baxter equation and unitarity relation. The standard equations for R-matrices are recovered by inserting the complete set of states provided by the stable basis $S^{(\pi)}_I(\vec\sigma)$ in between each elementary Janus interface. 



\section{Discussion} 
\label{sec:disc}

In this paper, we have investigated aspects of the correspondence between XXX$_{\frac{1}{2}}$ Heisenberg spin chains and 2d $\cN=(2,2)$ supersymmetric gauge theories. We have focussed on reproducing components of the algebraic Bethe ansatz for spin chains from correlation functions in $A$-twisted supersymmetric gauge theory and their reduction to partition functions in $\cN=4$ supersymmetric quantum mechanics. In particular, we have provided a concrete construction of the wavefunctions of off-shell Bethe states as orbifold defects in $A$-twisted supersymmetric gauge theory, and as thimble boundary conditions in supersymmetric quantum mechanics. We have also developed a new interpretation of the spin chain R-matrix as the matrix elements of Janus interfaces for mass parameters, leading to a novel and basis-independent presentation of the Yang-Baxter equations.   
 
Let us conclude with some directions for further research: 
 
\begin{itemize}
\item First, there are some important components of the algebraic Bethe ansatz that we have omitted from our presentation. One example is the generators of the Yangian symmetry of the spin chain. Unlike the R-matrix, Yangian generators have non-vanishing matrix elements between spin chain states with different number of excitations. On the supersymmetric side of the correspondence, this will correspond to correlation functions of interfaces that change the rank of the gauge group, $U(k)\to U(k')$. It is straightforward to construct such interfaces in the supersymmetric quantum mechanics description of section~\ref{sec:qm}, following methods introduced in~\cite{Bullimore:2016hdc}. However, we expect a complete discussion of Yangian representation theory and the algebraic Bethe ansatz will arise from `tri-partite' interfaces in supersymmetric quantum mechanics relating theories $(k,N)$, $(k',N')$ and $(k'',N'')$ with different gauge and flavour symmetries~\cite{BKL}. 
\item Secondly, in this paper we have considered only $\mathfrak{su}(2)$ spin chains with the fundamental representation at each site. It would be interesting to extend the results presented here to more general groups and representations, by studying more general supersymmetric quiver gauge theories. 
\item Finally, it would be interesting to extend our results to trigonometric or elliptic spin chains, corresponding to three and four dimensional supersymmetric gauge theories. The corresponding localization techniques for correlation functions of twisted theories on $S^2 \times S^1$ or $S^2\times T^2$ have been developed in~\cite{Benini:2015noa,Closset:2013sxa,Honda:2015yha}.
\end{itemize} 
  
\section*{Acknowledgements}
We would like to thank Stefano Cremonesi for useful discussions, and the organizers of the Pollica summer workshop for kind hospitality while this paper was being completed. MB and TL are supported by ERC STG grant 306260. HK would like to thank Mathematical Institute at University of Oxford and the 2017 Summer Workshop at the Simons Center for Geometry and Physics for their hospitality and support during different stages of this work. The research of HK is supported in part by NSF grant PHY-1067976.

 
\appendix

\section{Conventions} 
\label{app:conv}

\subsection{2d \texorpdfstring{$\cN=(2,2)$}{} Supersymmetry}

We consider two-dimensional $\mathcal{N}=(2,2)$ supersymmetric theory on a flat space with Euclidean coordinates $(x^1,x^2)$. We will also introduce a complex coordinate $z = x^1 + i x^2$. Our conventions are taken directly from Appendix A of~\cite{Herbst:2008jq} with $x^0 = - i x^2$.

The supersymmetry transformations of a vectormultiplet are
\begin{align}
\delta_\epsilon A_{\bar z} &=\frac{i}{2}(\bar\epsilon_+\lambda_+ + \epsilon_+\bar\lambda_+), \\
\delta_\epsilon A_z & = - \frac{i}{2}(\bar\epsilon_- \lambda_-+\epsilon_-\bar\lambda_-), \\
\delta_\epsilon\lambda_+ & =i\epsilon_+\left(D-2iF_{z\bar z} + \frac{1}{2}[\sigma,\bar\sigma] \right)+2 \epsilon_-D_{\bar z}\bar \sigma, \\
\delta_\epsilon\lambda_- & =i\epsilon_-\left(D+2iF_{z\bar z} - \frac{1}{2}[\sigma,\bar\sigma]  \right) - 2 \epsilon_+D_z\sigma, \\
\delta_\epsilon\bar\lambda_+ & =  - i\bar\epsilon_+\left(D+2iF_{z\bar z}+ \frac{1}{2}[\sigma,\bar\sigma]  \right)  +2 \bar \epsilon_-D_{\bar z} \sigma \\
\delta_\epsilon\bar\lambda_- & =   - i\bar\epsilon_-\left(D-2iF_{z\bar z} - \frac{1}{2}[\sigma,\bar\sigma]  \right)  - 2 \bar\epsilon_+ D_z\bar\sigma \\
\delta_\epsilon\sigma & = -i(\bar\epsilon_+\lambda_-+\epsilon_-\bar\lambda_+),\\ 
\delta_\epsilon\bar\sigma & = -i(\epsilon_+\bar\lambda_- + \bar\epsilon_-\lambda_+),\\ 
\delta_\epsilon D & = \bar\epsilon_+ D_z\lambda_+ - \bar\epsilon_- D_{\bar z} \lambda_- - \epsilon_+D_z \bar\lambda_+ + \epsilon_- D_{\bar z} \bar\lambda_-  \\
 & +\frac{i}{2}\epsilon_+[\sigma,\bar\lambda_-]+\frac{i}{2}\epsilon_-[\bar\sigma,\bar\lambda_+]-\frac{i}{2}\bar\epsilon_+[\bar\sigma,\lambda_-]-\frac{i}{2}\bar\epsilon_-[\sigma,\lambda_+].
\end{align}
For a chiral multiplet transforming in a unitary representation of the gauge group $G$, the supersymmetry transformations are
\begin{align}
\delta_\epsilon \phi & =\epsilon_+ \psi_--\epsilon_-\psi_+\,,\\
\delta_\epsilon \bar\phi & =-\bar\epsilon_+ \bar\psi_-+\bar\epsilon_-\bar\psi_+ \,,\\
\delta_\epsilon \psi_+ & = 2i\bar\epsilon_- D_{\bar z} \phi+\epsilon_+ F - \bar{\ep}_+ \bar \sigma  \phi  \,,\\
\delta_\epsilon \psi_- & = 2i\bar\epsilon_+ D_z \phi+\epsilon_- F + \bar\epsilon_- \sigma  \phi \,,\\
\delta_\epsilon \bar \psi_+ & = -2i\epsilon_- D_{\bar z} \bar\phi+\bar\epsilon_+ \bar F - {\ep}_+ \bar\phi\, \sigma    \,,\\
\delta_\epsilon \bar \psi_- & =- 2i\epsilon_+ D_z \bar\phi+\bar\epsilon_-\bar F + \epsilon_-\bar\phi\, \bar\sigma   \,, \\
\delta_{\epsilon}F & =-2i\bar \epsilon_-D_{\bar z} \psi_- + 2i\bar\epsilon_+D_z\psi_+ \\
& \quad + ( \bar \ep_+ \bar\sigma \psi_- + \bar \epsilon_- \sigma \psi_+) + i(\bar \epsilon_- \bar\lambda_+-\bar\epsilon_+\bar\lambda_-)\phi \,, \\
\delta_{\epsilon}\bar F & = -2i \epsilon_-D_{\bar z} \bar\psi_- + 2i\epsilon_+D_z\bar\psi_+ \\
& \quad - (  \ep_+  \bar\psi_-\sigma +  \epsilon_-  \bar\psi_+\bar\sigma) + i\bar\phi( \epsilon_- \lambda_+-\epsilon_+\lambda_-)\,.
\end{align}
where $D_\mu = \partial_\mu + i A_\mu$ and it is understood that vectormultiplet fields act in the appropriate representation of $G$.

The standard Yang-Mills Lagrangian for a vector multiplet is
\bea
\cL_V  =& \tr  \int d^4\theta \left(- \frac{1}{2e^2} \bar\Sigma \Sigma \right) \\
= &\frac{1}{2e^2}\tr\left( - 2 D_{\bar z} \sigma D_{z}\bar\sigma- 2 D_{z} \sigma D_{\bar z}\bar\sigma + 2 i \bar\lambda_- \overset{\leftrightarrow}{D_{\bar z}} \lambda_-  - 2 i \bar\lambda_+ \overset{\leftrightarrow}{D_{ z}} \lambda_+ + 4 F_{z\bar z}^2 + D^2\right.\\
&\left.  -\frac{1}{4}[\sigma,\bar \sigma]^2-\lambda_+[\sigma,\bar\lambda_-]-\lambda_-[\bar\sigma,\bar\lambda_+]-\bar\lambda_+[\bar\sigma,\lambda_-]-\bar\lambda_-[\sigma,\lambda_+]\right)
\eea
and FI term
\be
\cL_{FI} = \mathrm{Re} \int d^2\tilde\theta (- t \Sigma) = - r D - 2 \theta F_{z\bar z}
\ee
The chiral multiplet Lagrangian is
\bea
\cL & = \tr \int d^4 \theta \; \bar\Phi e^V \Phi + \mathrm{total\; derivative} \\
& = \tr\left(  -2 D_{\bar z}\phi D_z\bar \phi - 2 D_z\phi D_{\bar z} \bar\phi  +  i \bar\psi_- \overset{\leftrightarrow}{D_{\bar z}} \psi_-  - i \bar\psi_+ \overset{\leftrightarrow}{D_{z}} \psi_+
+ \bar\phi D \phi + |F|^2 - |\sigma \phi|^2  \right. \\ 
& \left. \quad - \bar\psi_-\sigma\psi_+ - \bar\psi_+ \bar\sigma \psi_- - i  \bar\phi \lambda_-\psi_+ + i  \bar\phi \lambda_+\psi_-  + i \bar\psi_+ \bar\lambda_-\phi - i \bar\psi_-\bar\lambda_+\phi +\frac{1}{2}\bar\phi[\sigma,\bar\sigma]\phi\right)
\eea
and superpotential term
\be
\cL_W = \mathrm{Re} \int d^2\theta W(\Phi)
\ee

Writing a general supersymmetry transformation as
\begin{equation}
\delta_\epsilon =i ( \epsilon_+ \Q_- -\epsilon_- \Q_+ -\bar{\epsilon}_+ \bar \Q_-   + \bar{\epsilon}_-\bar{\Q}_+ )
\end{equation}
we find that
\begin{align}
\{ \Q_-,\overline \Q_-\} & = 2 i D_z \, \quad && \{ \Q_+,\overline \Q_+\}  = -2 i D_{\bar z}  \\
\{ \Q_-,\overline \Q_+\} & = - \sigma  \quad && \{ \Q_+,\overline \Q_-\}  =   - \bar\sigma     \\
\qquad \{ \Q_+, \Q_-\} & =0 \quad  && \{ \overline \Q_+, \overline \Q_-\}=0 \\
 \Q_\pm^2 & = 0 \quad && \overline  \Q_\pm^2=0 \, . 
 \end{align}
The charges of the supersymmetry generators under $U(1)_J$ rotations and the axial $U(1)_A$ and vector $U(1)_V$ R-symmetries are shown below:
\begin{table}[h]
\begin{center}
\begin{tabular}{c|ccc|c}
& $U(1)_J$ & $U(1)_V$ & $U(1)_A$ & $U(1)_J'$ \\ \hline
$\Q_+$ & $-1$ & $-1$ & $-1$ & $-2$  \\
$\bar\Q_+$ & $-1$ & $+1$ & $+1$ & $0$ \\
$\Q_-$ & $+1$ & $-1$ & $+1$ & $0$ \\
$\bar\Q_-$ & $+1$ & $+1$ & $-1$ & $+2$
\end{tabular}
\end{center}
\end{table}
 
\subsection{\texorpdfstring{$\cN=4$}{} Quantum Mechanics}\label{app:SQM}
 
 In order to write the $\cN=(2,2)$ supersymmetry algebra as a 1d $\cN=4$ supersymmetry algebra, we compactify a spatial direction on a circle $x^1 \sim x^1 + 2\pi R$ and rename $\tau = x^2$. We can then organize the supercharges into spinors 
 \be
 \Q_\al = \begin{pmatrix}
 \Q_- \\ \Q_+
 \end{pmatrix} \qquad
 \bar \Q_{\al} = \begin{pmatrix}
 \bar \Q_+ \\ - \bar \Q_-
 \end{pmatrix}
 \ee
combining supercharges of $U(1)_V$ charge $-1$ and $+1$ respectively. Note that the top (resp. bottom) components of both spinors have charge $+1$ (resp.$-1$) under $U(1)_A$. With this notation, the supersymmetry algebra with $Z= 0 $ can be re-expressed as follows
\bea
\{ \Q_\al , \Q_{\beta} \} & = 0 \\
\{ \Q_\al , \bar \Q_{\beta} \} & = \ep_{\al\beta} D_\tau + Z_{\al\beta} \\
\{ \bar \Q_\al , \bar \Q_{\beta} \} & = 0
\eea
where
\be
\ep_{\al\beta} = \begin{pmatrix}
0 & -1 \\ 1 & 0
\end{pmatrix} \qquad
Z_{\al\beta} = \begin{pmatrix}
- \sigma &  - i D_1 \\ - i D_1 & \bar \sigma
\end{pmatrix}
\ee
This takes the form of an $\cN=4$ supersymmetric quantum mechanics with R-symmetry $U(1)_V \times SU(2)_A$ R-symmetry. States with KK momentum in the $x^1$ direction clearly break $SU(2)_A$ to the $U(1)_A$ axial R-symmetry in two dimensions.

We now write the supersymmetry transformations of the fields in $SU(2)_A$ covariant notation. We need to choose a convention for raising and lowering indices and will choose $\ep^{12}=\ep_{21}=1$ with $\psi^\al = \ep^{\al\beta} \psi_{\beta}$ and $\psi_\al = \ep_{\al\beta} \psi^{\beta}$.
We first write the supersymmetry transformations in $SU(2)_A$ covariant notation as
\be
\delta = i ( \ep^\al \Q_\al - \bar \epsilon^\al \bar \Q_{\al} )
\ee

For the vectormultiplet we write
\be
\sigma_{\al\beta} = \sigma^I (\tau^I)_{\alpha\beta}= \begin{pmatrix}
-\sigma &  A_1\\
  A_1 &  \bar\sigma
\end{pmatrix}
\ , \quad
\lambda_\al = \begin{pmatrix} \lambda_- \\ \lambda_+ \end{pmatrix}
\ , \quad \bar\lambda_\al = \begin{pmatrix} \bar \lambda_+ \\ -\bar \lambda_- \end{pmatrix} \ ,
\ee
where $(\tau^I)_\alpha^{\ \beta}$ are the Pauli matrices. In terms of these fields, the supersymmetry transformation is given by
\bea
	\delta A_\tau & = \frac{1}{2}(-\bar\epsilon^\alpha\lambda_\alpha+\epsilon^\alpha\bar\lambda_\alpha) \ , \quad \delta \sigma_{\alpha\beta} = i\epsilon_{(\alpha}\bar\lambda_{\beta)} + i\bar\epsilon_{(\alpha}\lambda_{\beta)}  \ , \\
	\delta \lambda_{\alpha} &= - i \epsilon^\beta D_{\tau}\sigma_{\alpha\beta} - \frac{i}{2}\epsilon_\gamma[\sigma_\alpha^{\ \beta},\sigma_{\beta}^{\ \gamma}]+ i\epsilon_\alpha D \ , \\
	\delta \bar\lambda_{\alpha} & = i \bar\epsilon^\beta D_\tau \sigma_{\alpha\beta}-\frac{i}{2}\bar\epsilon_\gamma[\sigma_\alpha^{\ \beta},\sigma_{\beta}^{\ \gamma}] - i\bar\epsilon_\alpha D \ , \\
	\delta D & = \frac{i}{2}\left(\bar\epsilon^\alpha D_\tau \lambda_\alpha + \epsilon^\alpha D_\tau \bar\lambda_\alpha +[\sigma_\alpha^{ \ \beta}, \bar\lambda_\beta]\epsilon^\alpha -[\sigma_{\alpha}^{\ \beta}, \lambda_\beta] \bar\epsilon^\alpha \right)\ .
\eea

For the chiral multiplet we have the supersymmetry transformation as
\bea
\delta \phi & = \ep^\al \psi_\al  \ ,
 \quad \delta\bar\phi = - \bar\epsilon^\alpha\bar\psi_\alpha \ ,
 \\
\delta \psi_\al & =   \bar \epsilon_\al D_\tau \phi + \ep_\al F + \bar\epsilon^\beta \sigma_{\al\beta} \phi \ , \\
\delta \bar\psi_\al &= \epsilon_\al D_\tau \bar \phi +\bar\ep_\al \bar{F}+\epsilon^\beta\bar\phi\,\sigma_{\al\beta} \ , \\
\delta F & = -\bar\epsilon^\al D_\tau \psi_\al - i \bar\epsilon^\al \bar \lambda_\al \phi +\bar\epsilon^\al \sigma_{\al\beta} \psi^\beta \ , \\
\delta\bar{F} &= \epsilon^\al D_\tau \bar\psi_\al - i\epsilon^\alpha\lambda_\alpha \bar\phi +\epsilon^\alpha \sigma_{\alpha\beta}\bar\psi^\beta \ ,
\eea
where
\be
\psi_\al = \begin{pmatrix} \psi_- \\ \psi_+ \end{pmatrix}
\ , \quad \bar\psi_\al = \begin{pmatrix} \bar \psi_+ \\ -\bar \psi_- \end{pmatrix} \ .
\ee

\bibliographystyle{JHEP}
\bibliography{Bethe_Gauge}

\end{document}